# Polaractivation of Hidden Private Classical Capacity Region of Quantum Channels

Laszlo Gyongyosi, *Member, IEEE*

[1]Quantum Technologies Laboratory, Department of Telecommunications
*Budapest University of Technology and Economics*
2 Magyar tudosok krt, Budapest, *H*-1117, Hungary
[2]Information Systems Research Group, Mathematics and Natural Sciences
*Hungarian Academy of Sciences*
Budapest, *H*-1518, Hungary

gyongyosi@hit.bme.hu

**Abstract**

We define a new phenomenon for communication over noisy quantum channels. The investigated solution is called polaractivation and based on quantum polar encoding. Polaractivation is a natural consequence of the channel polarization effect in quantum systems and makes possible to open the hidden capacity regions of a noisy quantum channel by using the idea of rate increment. While in case of a classical channel only the rate of classical communication can be increased, in case of a quantum channel the channel polarization and the rate improvement can be exploited to open unreachable capacity regions. We demonstrate the results for the opening of private classical capacity-domain. We prove that the method works for arbitrary quantum channels if a given criteria in the symmetric classical capacity is satisfied. We also derive a necessary lower bound on the rate of classical communication for the polaractivation of private classical capacity-domain.

# 1. Introduction

The polar coding technique was developed for classical systems to achieve the *symmetric* capacity of a classical noisy communication channel. The symmetric capacity is the highest rate at which the channel can be used for communication if the probability of the input letters is equal [1-9]. The channel polarization scheme introduced by Arikan [1] for classical channels is a revolutionary encoding and decoding scheme, which makes possible the construction of codewords to achieve the symmetric capacity. Recently, in the quantum setting, the polar coding scheme was studied by Wilde and Guha [16], by Renes *et al.* [17], by Wilde and Renes [24], [32], [43]. As was shown in [17] and [24] an efficient scheme also can be constructed for the quantum communication channels. The superactivation effect [13], [18], [36-38], [40-41] makes possible to use zero-capacity channels for communication. It was shown that it works for the quantum capacity [14][15], classical and quantum zero-error capacity [11][12], [10]. However the superactivation effect has several drawbacks, since it cannot be extended to the private classical capacity, and the effect is theoretically limited by the maps of channels and their initial capacities. In this paper we show that similar results can be obtained by a more general framework which is based on a completely different phenomenon. It requires only a special channel coding scheme and can be extended to the private classical capacity. Furthermore, it eliminates all of the main drawbacks of the superactivation effect. The proposed method does not depend on the maps of the channels and there is no need to combine different channels in a joint construction to achieve the positive capacity.

We show that the quantum polar coding can be used for the *polaractivation* of *hidden capacity regions* of a noisy quantum channel [26-28], [42]. Polaractivation opens those capacity-domains of a noisy quantum channel, which were initially not accessible due to noise of the channel. It is trivially not possible for a classical communication channel $N$, because classical channels can transmit only classical information. As follows for a classical channel only the *rate* of classical communication can be increased and no further capacity-domains can be reached. For a quantum channel, several other possibilities exist: it could transmit classical, private classical or quantum information, with completely different transmission characteristics and behavior. A quantum channel $\mathcal{N}$ which can transmit classical information has an accessible classical capacity-domain. However this channel could also have other *hidden* capacity-domains (for transmission of private classical or quantum information) which still remain inaccessible. By the proposed polaractivation effect these hidden capacity-domains of a noisy quantum channel can be opened. Our channel coding scheme can be used to transmit private classical information over channels that are so noisy that they cannot transmit any classical information *privately*. We present that the proposed polaractivation requires only the use of quantum polar encoding scheme and any quantum channel for which a given condition (see Section 2.5) in the rate of classical communication is satisfied can be used for private communication. By the proposed encoding scheme we open the private classical capacity region, which effect has its roots in the channel polarization effect and the rate of maximal achievable classical communication. The efficiency of the proposed polaractivation effect is $\mathcal{O}(n \log n)$, assuming $n$ uses of the channel.

This paper is organized as follows. In Section 2, we review the basic definitions of delivering private classical communication over a quantum channel. In Section 3 introduces the polar encoding scheme, while Section 4 discusses the proposed polaractivation scheme. In Section 5, we interpret our theorems and the proofs. In Section 6 we illustrate the theorems with numerical evidences. Finally, in Section 7, we conclude the results.

## 2. Preliminaries

In this section we overview the basic definitions and formulas related to the private classical communication over noisy quantum channels.

*2.1. The Quantum Channel*

The map of the quantum channel can be expressed with a special representation called the *Kraus Representation* [23], [39]. For a given input system $\rho_A$ and the quantum channel $\mathcal{N}$, this representation can be expressed as

$$\mathcal{N}(\rho_A) = \sum_i N_i \rho_A N_i^\dagger, \tag{1}$$

where $N_i$ are the Kraus operators, and $\sum_i N_i^\dagger N_i = I$ [23]. The *isometric extension* of $\mathcal{N}$ by means of the *Kraus Representation* can be expressed as

$$\begin{aligned}\mathcal{N}(\rho_A) &= \sum_i N_i \rho_A N_i^\dagger \to U_{A\to BE}(\rho_A)\\ &= \sum_i N_i \otimes |i\rangle_E.\end{aligned} \tag{2}$$

The action of the quantum channel $\mathcal{N}$ on an operator $|k\rangle\langle l|$, where $\{|k\rangle\}$ is an orthonormal basis, also can be given in operator form using the Kraus operator $N_{kl} = \mathcal{N}(|k\rangle\langle l|)$. By exploiting the property $UU^\dagger = P_{BE}$, for the input quantum system $\rho_A$:

$$\begin{aligned}U_{A\to BE}(\rho_A) &= U\rho_A U^\dagger\\ &= \left(\sum_i N_i \otimes |i\rangle_E\right) \rho_A \left(\sum_j N_j^\dagger \otimes \langle j|_E\right)\\ &= \sum_{i,j} N_i \rho_A N_j^\dagger \otimes |i\rangle\langle j|_E.\end{aligned} \tag{3}$$

If we trace out the environment, we get

$$Tr_E\left(U_{A\to BE}(\rho_A)\right) = \sum_i N_i \rho_A N_i^\dagger. \tag{4}$$

*2.2. The Classical Capacity*

The *classical capacity* $C(\mathcal{N})$ of a quantum channel $\mathcal{N}$ describes the maximum amount of classical information that can be transmitted through the channel. The *Holevo-Schumacher-Westmoreland* (HSW) theorem [33-34] defines this quantity for product state input (i.e. entanglement is not allowed in the input) as

$$\begin{aligned}C(\mathcal{N}) &= \max_{\text{all } p_i,\rho_i} \chi = \chi(\mathcal{N})\\ &= \max_{\text{all } p_i,\rho_i}\left[S\left(\mathcal{N}\left(\sum_i p_i \rho_i\right)\right) - \sum_i p_i S(\mathcal{N}(\rho_i))\right],\end{aligned} \tag{5}$$

where $S(\rho) = -Tr(\rho \log(\rho))$ is the von Neumann entropy, and $\chi$ is called the Holevo quantity and the maximum is taken over all $\{p_i,\rho_i\}$ ensembles of input quantum states [22],[23].

$\chi(\mathcal{N})$ is called the Holevo capacity of the channel $\mathcal{N}$. Hastings showed that the entangled inputs can increase the amount of received classical information [13], which resulted that

$$C(\mathcal{N}) = \lim_{n \to \infty} \frac{1}{n} \chi(\mathcal{N}^{\otimes n}), \qquad (6)$$

where $\mathcal{N}^{\otimes n}$ denotes the $n$ uses of the quantum channel $\mathcal{N}$.

*2.3. The Private Classical Capacity*

The *private classical capacity* $P(\mathcal{N})$ of quantum channel $\mathcal{N}$ describes the maximum rate at which the channel is able to send classical information through the channel between Alice (*A*) and Bob (*B*) in *secure* way i.e. without any information leaked about the plain text message to an malicious eavesdropper Eve (*E*) [31], [29-30].

The block diagram of a generic private quantum communication system is depicted in Fig. 1. The first output of the channel belongs to Bob and denoted by $\rho_B = \mathcal{N}(\rho_A)$ while the second "receiver" is the environment (i.e., the eavesdropper) *E*, with state $\rho_E = E(\rho_A)$. In Fig. 1, we also depict the encoding scheme. The phase carries the *data* and the amplitude is the *key* for the encryption i.e., Alice first encodes the phase (data) and then the amplitude (key). Bob applies it in the reverse order using his successive and coherent decoder, as was shown by Renes and Boileau in [17] [19]: he first decodes the *amplitude* (key) information in the *Z* basis. Then Bob continues the decoding with the *phase* information, in the *X* basis. (*For the detailed description see Section 4.*)

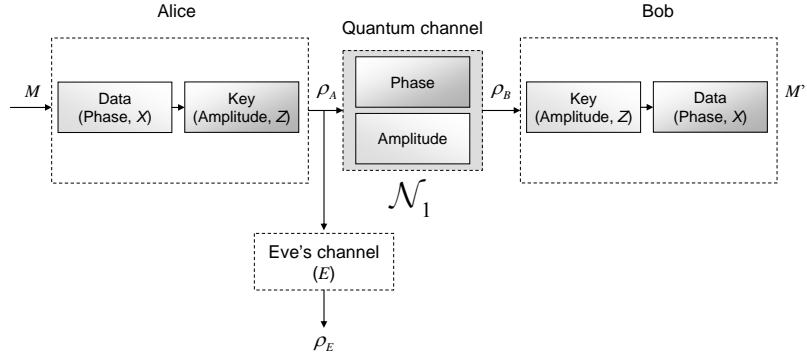

**Figure 1.** Private communication of Alice and Bob over a quantum channel in presence of an eavesdropper Eve. The quantum channel has positive private classical capacity if it can send both phase and amplitude.

The single-use (single-letter) private classical capacity (*private information*) can be expressed as the maximum of the difference between $I(A:B)$ which measures the classical information transmitted between Alice and Bob, and $I(A:E)$ that represents the information leaked to the eavesdropper [31]

$$P^{(1)}(\mathcal{N}) = \max_{\text{all } p_i, \rho_i} \left( I(A:B) - I(A:E) \right). \tag{7}$$

The optimization has to be taken over all possible source distributions and encoding schemes $\{p_i, \rho_i\}$ of Alice $\rho_A \in \{\rho_i\}$. The asymptotic private capacity can be defined as [31]

$$P(\mathcal{N}) = \lim_{n \to \infty} \frac{1}{n} P^{(1)}(\mathcal{N}^{\otimes n}) = \lim_{n \to \infty} \frac{1}{n} \max_{\text{all } p_i, \rho_i} \left( I(A:B) - I(A:E) \right). \tag{8}$$

The private capacity can be rewritten using the Holevo quantity as follows [31], [35]:

$$P(\mathcal{N}) = \lim_{n \to \infty} \frac{1}{n} \max_{\text{all } p_i, \rho_i} \left( \mathcal{X}_{AB} - \mathcal{X}_{AE} \right), \tag{9}$$

where

$$\mathcal{X}_{AB} = S(\mathcal{N}_{AB}(\rho_{AB})) - \sum_i p_i S(\mathcal{N}_{AB}(\rho_i)) \tag{10}$$

and

$$\mathcal{X}_{AE} = S(\mathcal{N}_{AE}(\rho_{AE})) - \sum_i p_i S(\mathcal{N}_{AE}(\rho_i)), \tag{11}$$

measure the Holevo quantities between Alice and Bob, and between Alice and Eve, respectively, where $\rho_{AB} = \sum_i p_i \rho_i$ and $\rho_{AE} = \sum_i p_i \rho_i$ are the average states. Alice's encoding transformation is amended with

$$\mathcal{E} : \{0,1\}^l \to \{0,1\}^n, \tag{12}$$

which takes the *l*-length input *M*, and from this message it constructs an *n*-length classical message before feeding transformation *X* and *Z*.

*2.4. The Symmetric Classical and Private Classical Capacity*

In our scheme, the recursive channel construction is the key ingredient to achieving the polarization effect, which splits the channels into two easily separable sets—one that contains zero-capacity channels, and one that contains nearly ideal channels. The fraction of the good channels converges the symmetric classical capacity. The *symmetric classical capacity* of the quantum channel is defined for *uniform* input distribution. An important property of the *symmetric* capacity, that there is *no maximization* in the mutual information function [16], [17] since the distribution of the input states is assumed to be *uniform*. Assuming the uniformly distributed $A$ classical input system with $a \in \{0,1\}$, the channel output system $B$ (consist of the channel output quantum states $\sigma_0, \sigma_1$) with respect to $A$ is defined by [16]

$$\sigma^{AB} = \frac{1}{2}|0\rangle\langle 0|^A \otimes \sigma_0 + \frac{1}{2}|1\rangle\langle 1|^A \otimes \sigma_1. \tag{13}$$

For a quantum channel $\mathcal{N}$ with input system $A$ and output system $B$, the *symmetric* classical capacity is equal to the symmetric quantum mutual information $I(A:B)$ [1], [16],

$$C_{sym}(\mathcal{N}) = I(A:B). \tag{14}$$

(*Note*: the $I(A:B)$ *symmetric quantum mutual information is additive for a quantum channel.*)

The result in (14) further can be evaluated as [6]

$$C_{sym}(\mathcal{N}) = \mathrm{S}\big((\sigma_0 + \sigma_1)/2\big) - \mathrm{S}(\sigma_0)/2 - \mathrm{S}(\sigma_1)/2. \tag{15}$$

For the *symmetric* private classical capacity [24], [32] the same condition holds, i.e., there is *no maximization* needed because the inputs are uniformly distributed and the channels between Alice and Bob, and Alice and Eve are symmetric:

$$P_{sym}^{(1)}(\mathcal{N}) = I(A:B) - I(A:E) \tag{16}$$

and

$$P_{sym}(\mathcal{N}) = \lim_{n \to \infty} \frac{1}{n}\big(I(A:B) - I(A:E)\big), \tag{17}$$

where $I(A:E)$ is the symmetric quantum mutual information function between Alice and Eve.

*2.5. A Required Condition on Polaractivation*

The scheme can be applied for any quantum channel $\mathcal{N}$ with symmetric classical capacity $C_{sym}^*(\mathcal{N})$, however there is a strict condition on the lower bound of the classical symmetric ca-pacity $C_{low}(\mathcal{N})$. The *initial* classical symmetric capacity (i.e., the maximal rate of private classical communication that can be achieved by any, non polar coding-based channel coding scheme) of the channel is denoted by $C_{sym}(\mathcal{N})$. Before this, we have to formalize two important connections between the classical capacities $C_{sym}(\mathcal{N})$, $C_{low}(\mathcal{N})$ and the private classical capacity, $P_{sym}(\mathcal{N})$ of the channel $\mathcal{N}$. (*Note: From now on, under the capacities we mean the maximal achievable transmission rates, see also Remarks 1 and 2.*)

**Proposition 1**. Assume the channel $\mathcal{N}$ with symmetric classical capacity $C^*_{sym}(\mathcal{N})$. If initially $C_{sym}(\mathcal{N}) < C_{low}(\mathcal{N}) \leq C^*_{sym}(\mathcal{N})$ holds, the quantum channel $\mathcal{N}$ cannot transmit any private classical information, thus $P_{sym}(\mathcal{N}) = 0$.

**Proposition 2**. The $P_{sym}(\mathcal{N}) > 0$ positive symmetric private capacity can be achieved if and only if $C_{sym}(\mathcal{N}) \geq C_{low}(\mathcal{N}) \leq C^*_{sym}(\mathcal{N})$.

From Propositions 1 and 2 follows, that the proposed quantum polar coding scheme can be used for private communication over any channel $\mathcal{N}$ for which the following condition holds:

$$C_{low}(\mathcal{N}) \leq C^*_{sym}(\mathcal{N}) \to P_{sym}(\mathcal{N}) > 0, \qquad (18)$$

i.e., if the $C_{low}(\mathcal{N}) \leq C^*_{sym}(\mathcal{N})$ critical lower bound on the symmetric classical capacity is exceeded the channel will have positive private capacity $P_{sym}(\mathcal{N}) > 0$. This condition is satisfied for all error-probabilities of the given quantum channel $\mathcal{N}$ for which the symmetric private capacity is positive for $C_{sym}(\mathcal{N}) \geq C_{low}(\mathcal{N}) \leq C^*_{sym}(\mathcal{N})$. The proposed scheme provides private communication over initially non-private quantum channels (i.e., over a channel $\mathcal{N}$ for which the private capacity-domain is *hidden*) if and only if $C^*_{sym}(\mathcal{N}) \geq C_{low}(\mathcal{N}) \geq C_{sym}(\mathcal{N})$. The goal of polaractivation is to achieve the following transformation between the classical symmetric capacities $C_{sym}(\mathcal{N})$ and $C_{low}(\mathcal{N})$:

$$C_{sym}(\mathcal{N}) < C_{low}(\mathcal{N}) \leq C^*_{sym}(\mathcal{N}) \to C_{sym}(\mathcal{N}) \geq C_{low}(\mathcal{N}) \leq C^*_{sym}(\mathcal{N}) \tag{19}$$

from which the desired transition

$$P_{sym}(\mathcal{N}) = 0 \to P_{sym}(\mathcal{N}) > 0 \tag{20}$$

follows.

**Remark 1.** *The improvement in the classical capacity means that we increase the rate $R$ of classical communication over the channel $\mathcal{N}$. Since the classical capacity $C^*_{sym}(\mathcal{N})$ of the channel is a "fixed" value, it cannot be modified. However, instead of this we will focus on the maximal rates $\max_{\forall \rho_i} R_{sym}(\mathcal{N})$, $\max_{\forall \rho_i} R_{low}(\mathcal{N})$ and $\max_{\forall \rho_i} R_{P_{sym}}(\mathcal{N})$ of the channel which are variable parameters. In the initial phase there is no exist input that could transmit private bits, but by the proposed polaractivation effect the maximized rate $\max_{\forall \rho_i} R_{sym}(\mathcal{N})$ can be increased above a critical limit $\max_{\forall \rho_i} R_{low}(\mathcal{N})$ which makes possible to open the private capacity domain, i.e., $\max_{\forall \rho_i} R_{P_{sym}}(\mathcal{N}) > 0$ will hold. Polaractivation increases the classical rate $\max_{\forall \rho_i} R_{sym}(\mathcal{N})$ by $\Delta$, where*

$$\max_{\forall \rho_i} R_{low}(\mathcal{N}) - \max_{\forall \rho_i} R_{sym}(\mathcal{N}) \leq \Delta \leq \max_{\forall \rho_i} R^*_{sym}(\mathcal{N}) - \max_{\forall \rho_i} R_{sym}(\mathcal{N}). \tag{21}$$

**Corollary 1.** *Based on Remark 1, the condition on the polaractivation defined in (19) can be rewritten as follows:*

$$\begin{aligned} \max_{\forall \rho_i} R_{sym}(\mathcal{N}) &< \max_{\forall \rho_i} R_{low}(\mathcal{N}) \leq \max_{\forall \rho_i} R^*_{sym}(\mathcal{N}) \to \\ \max_{\forall \rho_i} R_{sym}(\mathcal{N}) &\geq \max_{\forall \rho_i} R_{low}(\mathcal{N}) \leq \max_{\forall \rho_i} R^*_{sym}(\mathcal{N}) \end{aligned} \tag{22}$$

*from which we get*

$$\max_{\forall \rho_i} R_{P_{sym}}(\mathcal{N}) = 0 \to \max_{\forall \rho_i} R_{P_{sym}}(\mathcal{N}) > 0, \tag{23}$$

*where $\max_{\forall \rho_i} R_{sym}$, $\max_{\forall \rho_i} R_{low}$ and $\max_{\forall \rho_i} R_{P_{sym}}$ depict the maximal rates at which classical and private classical communication is possible over the channel. Since we are interested in the maximal* *rates at which classical and private classical information is possible, we will further refer on it by the capacities, as it is clarified in Definition 1.*

**Definition 1.** *The capacity formulas of $\mathcal{N}$ can be redefined by the maximal achievable rates as follows:*

$$C_{sym}^{*}\left(\mathcal{N}\right) \equiv \max_{\forall \rho_i} R_{sym}^{*}\left(\mathcal{N}\right), \tag{24}$$

$$C_{low}\left(\mathcal{N}\right) \equiv \max_{\forall \rho_i} R_{low}\left(\mathcal{N}\right), \tag{25}$$

$$C_{sym}\left(\mathcal{N}\right) \equiv \max_{\forall \rho_i} R_{sym}\left(\mathcal{N}\right) \tag{26}$$

and

$$P_{sym}\left(\mathcal{N}\right) \equiv \max_{\forall \rho_i} R_{P_{sym}}\left(\mathcal{N}\right). \tag{27}$$

Based on Definition 1, the maximized rates $\max_{\forall \rho_i} R_{sym}^{*}$, $\max_{\forall \rho_i} R_{sym}$, $\max_{\forall \rho_i} R_{low}$ and $\max_{\forall \rho_i} R_{P_{sym}}$ will be referred by the capacities $C_{sym}^{*}\left(\mathcal{N}\right)$, $C_{sym}\left(\mathcal{N}\right)$, $C_{low}\left(\mathcal{N}\right)$ and $P_{sym}\left(\mathcal{N}\right)$.

*2.5.1. Example for a Polaractivable Channel*

Next we demonstrate these statements with an exact quantum channel model. Our example is the *erasure* quantum channel $\mathcal{N}_p$, which erases the input state $\rho$ with probability $p$ or transmits the state with probability $(1-p)$:

$$\mathcal{N}_p\left(\rho\right) \rightarrow \left(1-p\right)\rho + \left(p|e\rangle\langle e|\right), \tag{28}$$

where $\rho$ is the output state and $|e\rangle$ is the erasure state. The classical capacity of the erasure quantum channel $\mathcal{N}_p$ can be expressed as

$$C_{sym}^{*}\left(\mathcal{N}_p\right) = \left(1-p\right)\log\left(d\right), \tag{29}$$

where $d$ is the dimension of the input system $\rho$. As follows from (29), the classical capacity of $\mathcal{N}_p$ vanishes at $p=1$, if $0 \leq p < 1$ then the channel $\mathcal{N}_p$ can transmit some classical information. The private capacity of the erasure quantum channel $\mathcal{N}_p$ is

$$P(\mathcal{N}_p) = (1-2p)\log(d). \tag{30}$$

The $P(\mathcal{N}_p)$ private capacity vanishes at $p = 1/2$, but it can transmit some private information if $0 \leq p < 1/2$.

In Fig. 2 the $C_{sym}(\mathcal{N}_p)$ initial symmetric classical (dashed line) and $P_{sym}(\mathcal{N}_p)$ symmetric private classical capacity (solid line) of the $\mathcal{N}_p$ erasure quantum channel as a function of erasure probability $p$ are shown. The channel has $C^*_{sym}(\mathcal{N}_p) = 0.55$, however initially it cannot transmit any private information, i.e., $C_{sym}(\mathcal{N}_p) < C_{low}(\mathcal{N}_p) \leq C^*_{sym}(\mathcal{N}_p)$. In this initial phase the $P_{sym}(\mathcal{N}_p)$ private capacity-domain of channel $\mathcal{N}_p$ is *hidden*. After the channels have being *polarized*, $C_{sym}(\mathcal{N}_p) \geq C_{low}(\mathcal{N}_p) \leq C^*_{sym}(\mathcal{N}_p)$ will hold, from which follows that $P_{sym}(\mathcal{N}_p) > 0$ is also satisfied. It requires the improvement of the rate of classical communication from $\max\limits_{\forall \rho_i} R_{sym}(\mathcal{N})$ to $\max\limits_{\forall \rho_i} R_{sym}(\mathcal{N}) + \Delta$. The symmetric capacity $C^*_{sym}(\mathcal{N}_p)$ of $\mathcal{N}_p$ is depicted by the thick green line.

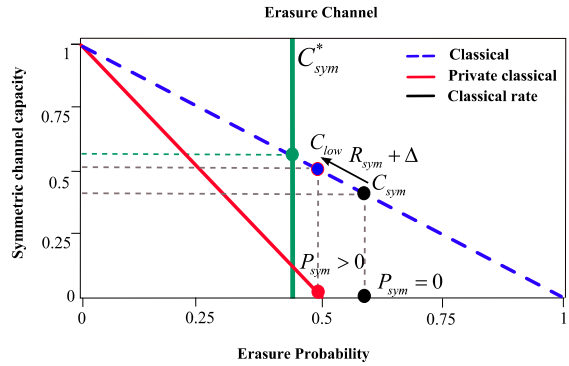

**Figure 2.** The polaractivation of private classical capacity domain is based on the idea of rate increment. The polaractivation works for any channels for which the condition $C_{low}(\mathcal{N}_p) \leq C^*_{sym}(\mathcal{N}_p) \rightarrow P_{sym}(\mathcal{N}_p) > 0$ is satisfied. The channel initially cannot transmit any private information, because $C_{sym}(\mathcal{N}_p) < C_{low}(\mathcal{N}_p) \leq C^*_{sym}(\mathcal{N}_p)$ holds, and the private classical capacity region of the channel is hidden. The capacities represent the maximal achievable transmission rates. The required improvement $\Delta$ in $C_{sym}(\mathcal{N}_p)$ is $C_{low}(\mathcal{N}_p) - C_{sym}(\mathcal{N}_p) \leq \Delta \leq C^*_{sym}(\mathcal{N}_p) - C_{sym}(\mathcal{N}_p)$. (These connections are illustrated with the erasure quantum channel $\mathcal{N}_p$ with $C^*_{sym}(\mathcal{N}_p) = 0.55$.)

**Corollary 2.** *The lower bound on the length l of transmittable private codewords is determined by* $C_{low}(\mathcal{N}_p)$.

**Corollary 3.** *The improvement in the rate* $\max_{\forall \rho_i} R_{sym}(\mathcal{N})$ *cannot be arbitrary high. The possible set of polaractivable channels is also determined by the distance* $\Delta$ *between the initial maximal classical rate and the critical lower bound:* $\Delta = C_{sym}(\mathcal{N}_p) - C_{low}(\mathcal{N}_p)$, *where* $C^*_{sym}(\mathcal{N}_p) \leq C_{low}(\mathcal{N}_p) > C_{sym}(\mathcal{N}_p)$. *According to (21), the distance* $\Delta$ *is restricted to the following range:* $C_{low}(\mathcal{N}_p) - C_{sym}(\mathcal{N}_p) \leq \Delta \leq C^*_{sym}(\mathcal{N}_p) - C_{sym}(\mathcal{N}_p)$.

*2.6. Degradable and Non-degradable Quantum Channel*

The main channel between Alice and Bob is denoted by $\mathcal{N}_{Bob}$, while Eve's channel is $\mathcal{N}_{Eve}$. Bob's channel is *degradable* if there exists a $\mathcal{D}$ *degradation channel,* which can be used by Bob to simulate Eve's channel, *i.e.,*

$$\mathcal{N}_{Eve} = \mathcal{D}\mathcal{N}_{Bob}. \tag{31}$$

For the error-probabilities of a *degradable* channel $\mathcal{N}_{Bob}$ (i.e., a *degraded* or *non-degradable* $\mathcal{N}_{Eve}$), the relation $p_{Eve} > p_{Bob}$ holds. For a *non-degradable* $\mathcal{N}_{Bob}$ (i.e., a *non-degraded* or *degradable* $\mathcal{N}_{Eve}$), $p_{Eve} < p_{Bob}$. In the proposed scheme it is assumed that Eve's channel is *symmetric*: it means that it both *degradable* and *non-degradable*; however, if Eve's channel is not symmetric, a prefix channel can be applied to have this property [21].

## 3. The Polar Encoding Scheme

Polar codes belong to the group of error-correcting codes. Polar coding was developed by Arikan for classical DMCs (*Discrete Memoryless Channels*) [1]. The polar codes introduce no redundancy only operate on codewords of $n$ of length, and can be used to achieve the symmetric capacity of classical DMCs. The basic idea behind the construction of polar codes is channel selection, called polarization: assuming $n$ identical DMCs we can create two sets of the logical channels by means of an encoder. "Good" channels are nearly noiseless while "bad" channels have nearly zero capacity. Furthermore, for large enough $n$, the fraction of good logical channels approaches the symmetric capacity of the original DMC. The polarization effect is represented by means of generator matrix $G_k$ having $k \times k$ of size [1] calculated in a recursive way

$$G_k = \left( I_{k/2} \otimes G_2 \right) R_k \left( I_2 \otimes G_{k/2} \right), \tag{32}$$

where

$$G_2 = \begin{pmatrix} 1 & 1 \\ 0 & 1 \end{pmatrix}, \tag{33}$$

and $I_{k/2}$ is the $\frac{k}{2} \times \frac{k}{2}$ identity matrix and $R_k$ is the $k \times k$ permutation operator, which permutes the input bits. Now we present how matrix $G$ is related to the polarization effect. For an input message $M$ having $n = 2^k$ length, the encoded codeword $A$ is $A = f(M) = G_k M$, i.e., if $k = 2$, then

$$G \begin{pmatrix} M_1 \\ M_2 \end{pmatrix} = \begin{pmatrix} M_1 \oplus M_2 \\ M_2 \end{pmatrix}. \tag{34}$$

For the transmission of an $n$-length encoded codeword $A$, and $k$-level recursion with matrix $G_k$, using $n$-times the classical channel $N$, the error-probability of the transmission of an $n$-length block is [1], [5]

$$p_{error}(A) = n 2^{-n\beta}. \tag{35}$$

The same results can be applied to a quantum channel $\mathcal{N}$, with classical input message $M$ and quantum output $\sigma$, and the XOR operation in (34) can implemented by a simple controlled-NOT (CNOT) gate. Assuming an $n$-length input codeword, the required number of CNOT gates is $\frac{1}{2} n \log_2 n$.

In Fig. 3(a) we show the first-level classical-quantum channel construction, where the input of the zero-level quantum channel $\mathcal{N}_1$ is the classical message $u_i$, while the output of $\mathcal{N}_1$ is the density matrix $\sigma_i$. The distinction of 'bad' channel $\mathcal{B}$, and in Fig. 3(b) the first-level, $\mathcal{N}_2$, 'good' channel $\mathcal{G}$ are also depicted [16]. The difference between the two channels is the knowledge of input $u_1$ on Bob's side. For the 'bad' channel $\mathcal{B}$ the input $u_1$ is unknown. In Fig. 3(c), the second-level channel $\mathcal{N}_4$ is shown, which is the combination of the two first-level channels $\mathcal{N}_2$. From the two quantum channels, a new one, denoted by $\mathcal{N}_2$, is constructed by a simple CNOT gate. The recursion can be repeated over $k$ levels, with $n = 2^k$ channel uses. The two independent $\mathcal{N}_2$ channels are combined into a higher-level channel, denoted by $\mathcal{N}_4$. The scheme also contains a permutation operator $R$, which permutes the control and target before the next level's CNOT gates in the channel structure. The outputs of $\mathcal{N}_4$ are the density matrices $\sigma_1, \sigma_2$.

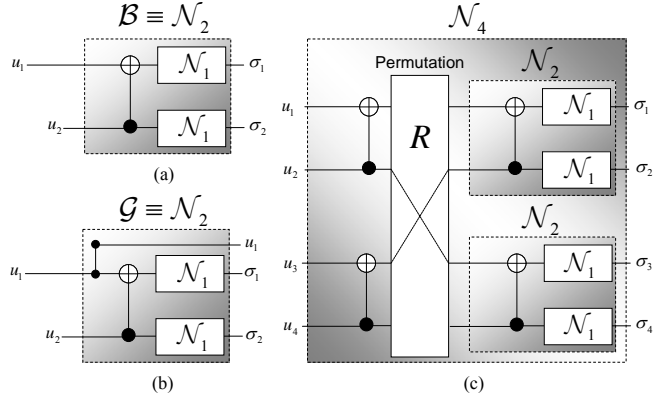

**Figure 3.** (a): The 'bad' channel: input $u_1$ is not known by Bob. (b): The 'good' logical channel: $u_1$ is also known on Bob's side. (c): The recursive channel construction from two lower-level channels. $R$ is the permutation operator. The channel structure has classical input and quantum output.

The 'bad' and 'good' channels from Fig. 3(a) and Fig. 3(b) are defined as follows [1,17]:

$$\begin{aligned}\mathcal{G} &\equiv \mathcal{N}\left(u_1, \sigma_1, \sigma_2 \middle| u_2\right) = \frac{1}{2}\mathcal{N}\left(\sigma_1 \middle| u_1 \oplus u_2\right)\mathcal{N}\left(\sigma_2 \middle| u_2\right), \\ \mathcal{B} &\equiv \mathcal{N}\left(\sigma_1, \sigma_2 \middle| u_1\right) = \frac{1}{2}\sum_{u_2 \in \{0,1\}}\mathcal{N}\left(\sigma_1 \middle| u_1 \oplus u_2\right)\mathcal{N}\left(\sigma_2 \middle| u_2\right).\end{aligned} \quad (36)$$

Using the polarized channel structure, for any $\beta < 0.5$, and assuming $n$ channel uses:

$$\lim_{k \to \infty} \left( \sqrt{F(\mathcal{N}_i)} < 2^{-2^{k\beta}} = 2^{-n^\beta} \right) = I(\mathcal{N}), \tag{37}$$

where $F$ is the fidelity of the $i$-th synthesized channel $\mathcal{N}_i$, and $k$ is the level of the encoder and decoder. The following important result can be derived from (36) for the mutual information of these channels for uniform inputs $u_1$ and $u_2$, namely [16]

$$\begin{aligned} &I(\mathcal{N}(\sigma_1, \sigma_2 | u_1)) + I(\mathcal{N}(u_1, \sigma_1, \sigma_2 | u_2)) \\ &= I(u_1 : \sigma_1 \sigma_2) + I(u_2 : u_1 \sigma_1 \sigma_2) \\ &= 2I(\mathcal{N}), \end{aligned} \tag{38}$$

where

$$I(u_1 : \sigma_1 \sigma_2) \leq I(\mathcal{N}) \tag{39}$$

and

$$I(u_2 : u_1 \sigma_1 \sigma_2) \geq I(\mathcal{N}), \tag{40}$$

i.e.,

$$\begin{aligned} I(u_1 : \sigma_1 \sigma_2) &\leq I(\mathcal{N}) \leq I(u_2 : u_1 \sigma_1 \sigma_2) \\ I(\mathcal{N}(\sigma_1, \sigma_2 | u_1)) &\leq I(\mathcal{N}) \leq I(\mathcal{N}(u_1, \sigma_1, \sigma_2 | u_2)). \end{aligned} \tag{41}$$

For the polarized 'bad' $\mathcal{B}(\mathcal{N}, \beta)$ and 'good' $\mathcal{G}(\mathcal{N}, \beta)$ channels the following rules hold:

$$\begin{aligned} \mathcal{G}(\mathcal{N}, \beta) &\equiv \left\{ i \in n : \sqrt{F(\mathcal{N}_i)} < 2^{-n^\beta} \right\}, \\ \mathcal{B}(\mathcal{N}, \beta) &= [n] \setminus \mathcal{G}(\mathcal{N}_i, \beta), \end{aligned} \tag{42}$$

where $n = 2^k$. In (42), parameter $\beta$ is defined as

$$\beta = \frac{1}{n} \sum_{i=1}^{n} \log_n d_i, \tag{43}$$

where $d_i = d_{\min}(\mathbf{g_i}, \mathbf{g_{i+1}}, \ldots, \mathbf{g_n})$, and $g_i$ is the $i^{\text{th}}$ row vector of matrix $G_n$ As was shown by Arikan [1], Korada, Sasoglu, and Urbanke [6], $\beta \leq 0.5$ if $k < 15$, while for $k \geq 16$: $\beta > 0.5$, along with

$$\lim_{n\to\infty} \frac{1}{n}\beta = 1. \tag{44}$$

Private classical communication over these structures means the following: in her message $A$, Alice sends her encoded private message $M$ only over channels $\mathcal{G}(\mathcal{N},\beta)$, while the remaining parts of $A$ are transmitted via $\mathcal{B}(\mathcal{N},\beta)$. Moreover, after the channels are have being polarized, the fraction of $\mathcal{G}(\mathcal{N},\beta)$ will be equal to the symmetric private classical capacity $P_{sym}(\mathcal{N})$. For the private communication scenario (see Fig 1.) the polarized channel construction $\mathcal{N}^{\otimes n}$, assuming a sufficient high $n$ and $\beta < 0.5$, the following relation holds for a given set of codewords:

$$\begin{aligned} S_{Bob}(\mathcal{N}_{Bob},\beta) &\equiv \left\{ i \in n : \sqrt{F(\mathcal{N}_{i,Bob})} < 2^{-n^{\beta}} \right\}, \\ S_{Eve}(\mathcal{N}_{Eve},\beta) &\equiv \left\{ i \in n : \sqrt{F(\mathcal{N}_{i,Eve})} \geq 1 - 2^{-n^{\beta}} \right\}. \end{aligned} \tag{45}$$

For these codewords, the $P_{sym}$ achievable symmetric private classical communication capacity assuming non-symmetric channel $\mathcal{N}_{Eve}$ (i.e., in this case *maximization is required*) can be expressed as the difference of the two mutual information functions,

$$P_{sym}(\mathcal{N}) = \lim_{n\to\infty} \frac{1}{n} \max_{\text{all } p_i,\rho_i} \left( I(A:B) - I(A:E) \right). \tag{46}$$

*3.1. The Fidelity Parameter and Likelihood Ratio*

In the description of the polarization effect of a classical channel $N$, the Bhattacharya parameter $B(N)$ was used to describe the noise of the transmission [1]. For a quantum channel $\mathcal{N}$ it is analogous to the fidelity $F$ of the transmission [16], [17], which can be defined for mixed channel output states $\sigma$ and $\rho$ as follows

$$F(\rho,\sigma) = \left[ Tr\left(\sqrt{\sqrt{\sigma}\rho\sqrt{\sigma}}\right) \right]^2 = \sum_i p_i \left[ Tr\left(\sqrt{\sqrt{\sigma_i}\rho_i\sqrt{\sigma_i}}\right) \right]^2. \tag{47}$$

*Note: the Bhattacharya parameter B also could be used in the quantum setting since it also describes the amount of noise on the channel. However, we use the fidelity parameter F since it is*

*unambiguously related to quantum channels in comparison to the Bhattacharya parameter; however the role of the two parameters is exactly the same.*

Next we list the major properties of fidelity:

$$0 \leq F(\sigma,\rho) \leq 1, \tag{48}$$

$$F(\sigma,\rho) = F(\rho,\sigma), \tag{49}$$

$$F(\rho_1 \otimes \rho_2, \sigma_1 \otimes \sigma_2) = F(\rho_1,\sigma_1)F(\rho_2,\sigma_2), \tag{50}$$

$$F(U\rho U^\dagger, U\sigma U^\dagger) = F(\rho,\sigma), \tag{51}$$

$$F(\rho, a\sigma_1 + (1-a)\sigma_2) \geq aF(\rho,\sigma_1) + (1-a)F(\rho,\sigma_2), \ a \in [0,1]. \tag{52}$$

If Alice chooses an *l*-length message $M$ with *uniform* probability distribution, then the probability that Bob's decoder will not fail

$$p(M' = M) \geq 1 - \sum_{i=1}^{n} F(\mathcal{N}_i), \tag{53}$$

Assuming that Alice's *l*-length message $M$ is selected uniformly from a set of size $K$ and transmitted by means of an *n*-length codeword, then the reliability of the transmission can be expressed by the average error probability $p(M' \neq M)$, as follows:

$$p(M' \neq M) = \frac{1}{K}\sum_{k=1}^{K} \Pr(M' \neq k | M = k), \tag{54}$$

where

$$\lim_{n \to \infty} p(M' \neq M) = 0. \tag{55}$$

Using a *k*-level structure, for a given set of channels indices $F(\mathcal{N}_i) = 0$ or $F(\mathcal{N}_i) = 1$, respectively. If a synthesized logical channel is "good," $\mathcal{G}(\mathcal{N},\beta)$, then it can transmit perfectly the input, thus

$$F(\mathcal{N}_i) \to 0, \tag{56}$$

while if the synthesized logical channel belongs to the "bad" set, $\mathcal{B}(\mathcal{N},\beta)$, then

$$F(\mathcal{N}_i) \to 1. \tag{57}$$

The two different values of $F(\mathcal{N}_i)$ indicate that for large enough uses $n$ of the quantum channel structure $\mathcal{N}$, the channel structure $\mathcal{N}^{\otimes n}$ will be polarized. For the channels from set $\mathcal{G}(\mathcal{N},\beta)$, $F(\mathcal{N}_i) \to 0$ while the capacity of the channels will be nearly 1. For the channels from set $\mathcal{B}(\mathcal{N},\beta)$, $F(\mathcal{N}_i) \to 1$, i.e., the capacity of these channels will be nearly equal to 0. In the polar encoding, for the $\mathcal{B}(\mathcal{N},\beta)$ channels, Alice freezes the inputs, and valuable information will be transmitted only over the $\mathcal{G}(\mathcal{N},\beta)$ channels. The freezing can be made by choosing a determined input to $\mathcal{B}(\mathcal{N},\beta)$, for which the value is also known for Bob on the decoding side [1], [16]. The convergence of the fidelity parameters of the $\mathcal{G}(\mathcal{N},\beta)$ "good" and $\mathcal{B}(\mathcal{N},\beta)$ "bad" channel sets, and the steps of the iteration process was proven by Arikan and Telatar in [4] and by Arikan [1], [2]. In the description of the convergence of the iteration, another important parameter was also introduced, namely the $L$ likelihood ratio, which is used for the description of Bob's decoding strategy. The $L$ likelihood parameter measures whether the original input on Alice side was 0 or 1. From this value, Bob can decide with high probability, whether Alice sent 0 or 1. The likelihood parameter-based iteration is optimal and can be achieved in $\mathcal{O}(n \log n)$ time. For the quantum channel $\mathcal{N}$ with a given $n$-length output density matrix $\sigma$, the likelihood ratio [1] of the channel is defined as

$$L(\mathcal{N},\sigma) = \frac{\mathcal{N}(\sigma|0)}{\mathcal{N}(\sigma|1)}. \tag{58}$$

The likelihood parameter is computed for each of the $n$ bits of the received codeword $\sigma$. The $L(\mathcal{N},y)$ can be defined in a different way for the sets $\mathcal{G}(\mathcal{N},\beta)$ and $\mathcal{B}(\mathcal{N},\beta)$, since for the "bad" channels (see Fig. 3(b)), it can be expressed as

$$L\left(\mathcal{B}\left(\mathcal{N},\beta\right),\sigma_1,\sigma_2\right) = \frac{1+L\left(\mathcal{N},\sigma_1\right)L\left(\mathcal{N},\sigma_2\right)}{L\left(\mathcal{N},\sigma_1\right)+L\left(\mathcal{N},\sigma_2\right)}$$
$$= \frac{1+\left(\frac{\mathcal{N}(\sigma_1|0)}{\mathcal{N}(\sigma_1|1)}\right)\left(\frac{\mathcal{N}(\sigma_2|0)}{\mathcal{N}(\sigma_2|1)}\right)}{\frac{\mathcal{N}(\sigma_1|0)}{\mathcal{N}(\sigma_1|1)}+\frac{\mathcal{N}(\sigma_2|0)}{\mathcal{N}(\sigma_2|1)}}. \quad (59)$$

For the set $\mathcal{G}\left(\mathcal{N},\beta\right)$, we have to distinguish the case input $u_1 = 0$ and $u_1 = 1$, since for the "good" channels, $u_1$ is also known at Bob's side (see Fig. 3(b)). For $\mathcal{G}\left(\mathcal{N},\beta\right)$ with $u_1 = 0$,

$$L\left(\mathcal{G}\left(\mathcal{N},\beta\right),\sigma_1,\sigma_2\right) = L\left(\mathcal{N},\sigma_1\right)L\left(\mathcal{N},\sigma_2\right)$$
$$= \frac{\mathcal{N}(\sigma_1|0)}{\mathcal{N}(\sigma_1|1)}\frac{\mathcal{N}(\sigma_2|0)}{\mathcal{N}(\sigma_2|1)}, \quad (60)$$

while for $u_1 = 1$,

$$L\left(\mathcal{G}\left(\mathcal{N},\beta\right),\sigma_1,\sigma_2\right) = \frac{L\left(\mathcal{N},\sigma_2\right)}{L\left(\mathcal{N},\sigma_1\right)}$$
$$= \frac{\mathcal{N}(\sigma_2|0)}{\mathcal{N}(\sigma_2|1)}\frac{\mathcal{N}(\sigma_1|1)}{\mathcal{N}(\sigma_1|0)}. \quad (61)$$

Depending on the value of $L$, Bob will decide as follows. If $L > 0$, then he will decide for 0, since in this case 0 is more likely than 1. If $L < 0$, Bob will decide for 1.

## 4. The Polaractivation Encoding and Decoding Scheme

**Method.** *The polaractivation of private classical capacity of arbitrary quantum channels (for which the requirements of Section 2.5 are satisfied) requires only the proposed polar encoding scheme and the multiple uses of the same quantum channel.*

The quantum channel $\mathcal{N}$ has some positive symmetric capacity $C_{sym} > 0$, while it cannot transmit private information, since $C_{sym}\left(\mathcal{N}\right) < C_{low}\left(\mathcal{N}\right) \to P_{sym}\left(\mathcal{N}\right) = 0$ (for illustration see Fig. 2). We prove that the required transition $C_{sym}\left(\mathcal{N}\right) \geq C_{low}\left(\mathcal{N}\right) \to P_{sym}\left(\mathcal{N}\right) > 0$ can be achieved.

*4.1. Encoding of Private Classical Information*

The proposed polar coding scheme assumes the use of noisy quantum channels with amplitude and phase coding, similar to the scheme of Renes *et al.* [17]. The parties can use either the amplitude or the phase to encode classical information; however, the transmission of private classical *information* requires both amplitude and phase coding *simultaneously*. If Alice wants to send Bob classical (i.e., not private) information, then she can encode her information either into the amplitude or phase using the $Z$ and $X$ bases. It is possible for quantum channels, since for these channels the polarization occurs in *both* amplitude and phase [17]. If she wants to send her classical information privately, then she has to encode her information *both* in the amplitude (in the $Z$ basis) and in the phase (in the $X$ basis). As shown by Christandl and Winter [20], if Alice can send both amplitude and phase, then she can also send entanglement to Bob.

In Fig. 4, we depict our encoding scheme. The phase carries the *data*, while the amplitude is the *key* for the encryption i.e., in our scheme Alice first encodes the phase (data) and then the amplitude (key). In the decoding, Bob uses his successive and coherent decoder [17, 19]: he first decodes the *amplitude* (key) information in the $Z$ basis. Then Bob continues the decoding with the *phase* information, in the $X$ basis. The successful decoding of the amplitude information (key) is a necessary but not sufficient condition for the positive symmetric private classical capacity $P_{sym}(\mathcal{N})$.

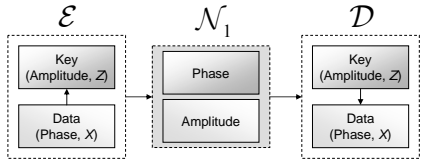

**Figure 4.** The encoding and decoding of private classical information (the zero-th level of recursion). To send classical information privately, the channel has to be able to send both amplitude and phase information at the same time. Alice encodes her private classical information into the amplitude and phase simultaneously.

According to our encoding scheme, the *symmetric private classical capacity $P_{sym}$* of $\mathcal{N}$ is defined as

$$P_{sym}(\mathcal{N}) = \lim_{n\to\infty} \frac{1}{n}(I(A:B) - I(A:E))$$
$$= \lim_{n\to\infty} \frac{1}{n} \begin{bmatrix} S((\sigma_0^{phase} + \sigma_1^{phase})/2) - S(\sigma_0^{phase})/2 \\ -S(\sigma_1^{phase})/2 - I(A:E) \end{bmatrix}, \quad (62)$$

where $I(A:E) = S(A) + S(E) - S(AE)$ stands for the mutual information function between Alice and Eve [6], [13], [21]. To construct the input polar codeword sets, we use the notation of 'good' $\mathcal{G}(\mathcal{N}, \beta)$ and 'bad' $\mathcal{B}(\mathcal{N}, \beta)$, where $\beta < 0.5$ [3], [6], [13], [21], channels for the transmission of phase and amplitude.

*4.2. Detailed Description*

In this section we show that the proposed polar encoding scheme can be applied for the *polaractivation* of the private classical capacity of arbitrary channels (for which the conditions of Section 2.5 are satisfied).

Let us assume that Alice has the $d$-dimensional classically correlated input system $\rho_A = |\Phi\rangle\langle\Phi|_A$ [17], [19-21]. In the encoding process, she encodes it with her encoding operator $U_\mathcal{E}$, and then sends through $\mathcal{N}$ to Bob. Alice has $n$ states, and she encodes the phase and the amplitude into $\rho_A = |\Phi\rangle\langle\Phi|_A$ producing $\tau_A = |\varphi\rangle\langle\varphi|_A$. Alice's encoder $\mathcal{E}$ realizes an $U_\mathcal{E}$ encoding transformation by using the $X$ and $Z$ operators for the encoding of the data and key information, which encoding process can be described as follows:

$$\mathcal{E}: |\varphi\rangle_A^{\otimes n} = U_\mathcal{E}|\Phi\rangle_A^{\otimes n} = \frac{1}{d^n} \sum_{key,data \in \{0,1\}^n} (-1)^{key \cdot G_{d^n} data} |data, key\rangle_A$$
$$= \frac{1}{d^n} \sum_{key,data \in \{0,1\}^n} (-1)^{G_{d^n}^T key \cdot data} |data, key\rangle_A, \quad (63)$$

where $G$ is the generator matrix $G = \begin{pmatrix} 1 & 1 \\ 0 & 1 \end{pmatrix}$, while $G^T$ is the transpose of $G$, and

$$|key\rangle \cdot G_{d^n} |data\rangle = G_{d^n}^T |key, data\rangle$$
$$= G_{d^n}^T \left( \frac{1}{d^n} \sum_{data \in \{0,1\}^n} (-1)^{key \cdot data} |key\rangle \right) \cdot |data\rangle. \quad (64)$$

In the basis of the data, $|\varphi\rangle_A^{\otimes n}$ can be expressed as:

$$|\varphi\rangle_A^{\otimes n} = \frac{1}{\sqrt{d^n}} \sum_{data \in \{0,1\}^n} |data, G_{d^n} data\rangle_A, \qquad (65)$$

while in the basis of the key, the state $|\varphi\rangle_A^{\otimes n}$ is:

$$|\varphi\rangle_A^{\otimes n} = \frac{1}{\sqrt{d^n}} \sum_{key \in \{0,1\}^n} |key, G_{d^n}^T key\rangle_A. \qquad (66)$$

The encoded system $|\varphi\rangle_A^{\otimes n}$ will be sent through the noisy quantum channel $\mathcal{N}$, which will transform $|\varphi\rangle_A^{\otimes n}$ into the output system $|\sigma\rangle_{BE}^{\otimes n}$ [17], [24]:

$$\begin{aligned}
\mathcal{N}|\varphi\rangle\langle\varphi|_A^{\otimes n} &= |\sigma\rangle_{BE}^{\otimes n} = U_\mathcal{N}^{BE} |\varphi\rangle_A^{\otimes n} |0\rangle_E \\
&= \sum_{data, key \in \{0,1\}^n} \sqrt{p_{data,key}} \left( X^{data} Z^{key} \right)_B |data, key\rangle_E,
\end{aligned} \qquad (67)$$

where $U_\mathcal{N}^{BE}$ denotes the transformation of channel $\mathcal{N}_{BE}$ between Bob and the $E$ environment, $|data, key\rangle$ are the environment states, and $p_{data,key}$ is the error probability of the transmission (i.e., the probability of a bit-flip of phase-flip error). In the decoding process, Bob with his $U_\mathcal{D} = U_\mathcal{E}^\dagger$ decoder transformation first decodes the key in basis $Z$ and then the data in basis $X$ of $|\sigma\rangle_{BE}^{\otimes n}$, which will result in system output state [17], [24]:

$$\begin{aligned}
|\varphi\rangle_B^{\otimes n} &= U_\mathcal{D}\left(|\sigma\rangle_{BE}^{\otimes n}\right) = U_\mathcal{E}^\dagger \left( U_\mathcal{N}^{BE} |\varphi\rangle_A^{\otimes n} |0\rangle_E \right) \\
&= U_\mathcal{E}^\dagger \left( \sum_{data, key \in \{0,1\}^n} \sqrt{p_{data,key}} \left( X^{data} Z^{key} \right)_B |\varphi\rangle_A |data, key\rangle_E \right) \\
&= \frac{1}{\sqrt{d^n}} \sum_{data, key \in \{0,1\}^n} \sqrt{p_{data,key}} |data, key\rangle_B |data, key\rangle_E.
\end{aligned} \qquad (68)$$

**Remark 2**. *An important notation has to be made on the polaractivation of quantum capacity-domain. From the encoding scheme of Section 4.2 follows that the quantum capacity-domain of the channel also can be opened if Alice uses a d-dimensional entangled system $\rho_{AB} = |\Phi\rangle\langle\Phi|_{AB}$* *for the encoding. The encoding operation $U_\mathcal{E}$ will act on the entangled system AB and subsystem B will be sent to Bob. The process of the encoding and decoding is nearly the same as was demonstrated for the polaractivation of the private classical capacity region.*

## 5. Theorems and Proofs

In this section we present the theorems and the proofs regarding the proposed coding scheme.

**Theorem 1.** *The polaractivation of the symmetric private classical capacity of arbitrary quantum channels (for which the conditions of Section 2.5 are satisfied) results in a non-empty set of polar codewords which set achieves the symmetric private classical capacity of the quantum channel.*

*Proof.* First we construct the input codewords and show that while initially the set of polar codewords which can transmit private classical information is empty in the initial phase, by the proposed polar encoder this set can be transformed into a non-empty set.

The codeword set $\mathcal{C}$ for *classical* (non-private) communication between Alice and Bob is defined as follows:

$$\mathcal{C} = \mathcal{G}(\mathcal{N}_{amp}, \beta), \tag{69}$$

where $|\mathcal{C}| = n$. The $S_{in}$ set of polar codewords that can transmit private information is defined as follows [17], [24], [32]:

$$S_{in} = \mathcal{G}(\mathcal{N}_{amp}, \beta) \cap \mathcal{G}(\mathcal{N}_{phase}, \beta), \tag{70}$$

where $|S_{in}| = l$. All of the other input codewords cannot be used for private classical communication and defined by the set $S_{bad}$ as follows

$$\begin{aligned} S_{bad} = &\left(\mathcal{G}(\mathcal{N}_{amp}, \beta) \cap \mathcal{B}(\mathcal{N}_{phase}, \beta)\right) \\ &\cup \left(\mathcal{B}(\mathcal{N}_{amp}, \beta) \cap \mathcal{G}(\mathcal{N}_{phase}, \beta)\right) \\ &\cup \left(\mathcal{B}(\mathcal{N}_{amp}, \beta) \cap \mathcal{B}(\mathcal{N}_{phase}, \beta)\right), \end{aligned} \tag{71}$$

where $|S_{bad}| = n - l$. Choosing $n = |\mathcal{G}(\mathcal{N}_{amp}, \beta)|$, $|S_{bad}|$ is equal to

$$|S_{bad}| = |\mathcal{B}(\mathcal{N}_{amp}, \beta) \cap \mathcal{G}(\mathcal{N}_{phase}, \beta)|. \quad (72)$$

From set $S_{bad}$, we define the completely useless codewords as [17], [24], [32]

$$\mathcal{B} = \mathcal{B}(\mathcal{N}_{amp}, \beta) \cap \mathcal{B}(\mathcal{N}_{phase}, \beta), \quad (73)$$

while the 'partly good' (i.e., can be used for non-private classical communication) input codewords will be denoted by [17], [24], [32]

$$\mathcal{P}_1 = \mathcal{G}(\mathcal{N}_{amp}, \beta) \cap \mathcal{B}(\mathcal{N}_{phase}, \beta) \quad (74)$$

and

$$\mathcal{P}_2 = \mathcal{B}(\mathcal{N}_{amp}, \beta) \cap \mathcal{G}(\mathcal{N}_{phase}, \beta). \quad (75)$$

The input codewords from $\mathcal{P}_1$ and $\mathcal{P}_2$ cannot be used to transmit classical information *privately*, since these codewords do not satisfy our requirements on the encoding of private information (only set $S_{in}$ is allowed in the proposed scheme). For a degradable channel with $n = |\mathcal{G}(\mathcal{N}_{amp}, \beta)|$, $|S_{in}| = l$, leads to $|S_{bad}| = n - l - b$, where $b = 0$. For a non-degradable channel, using $n = |\mathcal{G}(\mathcal{N}_{amp}, \beta)| + |\mathcal{B}|$, $|S_{in}| = l$ leads to relation $|S_{bad}| = n - l - b$, $b = |\mathcal{B}|$. The sets of the codewords are shown in Fig. 5.

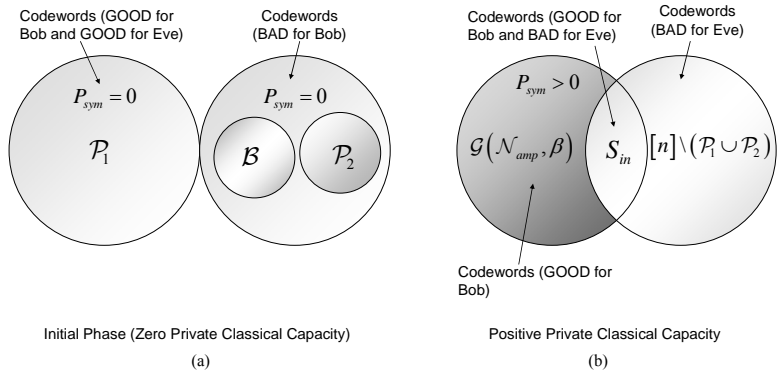

Initial Phase (Zero Private Classical Capacity)
(a)

Positive Private Classical Capacity
(b)

**Figure 5 (a).** The set of the constructed codewords for the transmission of private information cannot be accessed before the polaractivation. **(b)**. Due to the improvement in $n$ and the length ($n$-$l$) of classical codewords, the transmission of $l$-length private codewords will be possible.

Before the polaractivation is realized, the quantum channel $\mathcal{N}$ could not transmit any private classical information, i.e.:

$$S_{in} = \mathcal{G}(\mathcal{N}_{amp}, \beta) \cap \mathcal{G}(\mathcal{N}_{phase}, \beta) = \varnothing, \qquad (76)$$

and $|S_{in}| = 0$.

**Proposition 3.** (On the rate of classical communication). *The phenomenon of polaractivation of private classical capacity region based on the following fact: by exploiting the polarization effect, the rate of classical, non-private communication can be increased. We have to select those codeword sets that can transmit classical information, then the length n of these codewords have to be increased. As result, the required improvement in the rate of classical communication can be achieved and the private classical capacity region will become available for the transmission of l-length private codewords.*

*Note: the length of the codewords represents the number of channel indices of good logical channels. For set $\mathcal{C}$ these are the good channels that can transmit amplitude information. For the set $S_{in}$, these logical channels can transmit both amplitude and phase.*

First we have to define the codewords set $\mathcal{C}$, that is able to transmit classical information between Alice and Bob. This codeword sets can be defined as follows:

$$\begin{aligned}
\mathcal{C} &= \mathcal{P}_1 \cup S_{in} \\
&= \left(\mathcal{G}(\mathcal{N}_{amp}, \beta) \cap \mathcal{B}(\mathcal{N}_{phase}, \beta)\right) \cup \left(\mathcal{G}(\mathcal{N}_{amp}, \beta) \cap \mathcal{G}(\mathcal{N}_{phase}, \beta)\right) \\
&= \left(\mathcal{G}(\mathcal{N}_{amp}, \beta) \cup \mathcal{G}(\mathcal{N}_{amp}, \beta)\right) \cap \left(\mathcal{B}(\mathcal{N}_{phase}, \beta) \cup \mathcal{G}(\mathcal{N}_{phase}, \beta)\right) \\
&= \mathcal{G}(\mathcal{N}_{amp}, \beta).
\end{aligned} \qquad (77)$$

From (77) follows, that the codewords that can be used to increase the rate of classical communication $C_{sym}(\mathcal{N}) = \max_{\forall \rho_i} R_{sym}(\mathcal{N})$ to above the critical lower bound $C_{low}(\mathcal{N}) = \max_{\forall \rho_i} R_{low}(\mathcal{N})$, see (26) and (25), are those codewords that can be used to send amplitude information from Alice to Bob, i.e., $\mathcal{C} = \mathcal{G}(\mathcal{N}_{amp}, \beta)$. The required condition on the rate of classical communication to open the private classical capacity region (see (22)) in terms of the codeword length (*n-l*) can be expressed as follows:

$$|\mathcal{C}| = |\mathcal{G}(\mathcal{N}_{amp}, \beta)| \rightarrow |\mathcal{G}(\mathcal{N}_{amp}, \beta)| > |\mathcal{C}_{low}|, \tag{78}$$

where $|\mathcal{C}_{low}|$ is the critical lower bound on the number of channel indexes in the polarized structure (in other words the length on these codewords) that transmit classical information, from which the transition in the codeword set

$$S_{in} = \varnothing \rightarrow S_{in} \neq \varnothing \tag{79}$$

can be realized. These conditions in terms of the polar codeword sets are summarized in Fig. 6.

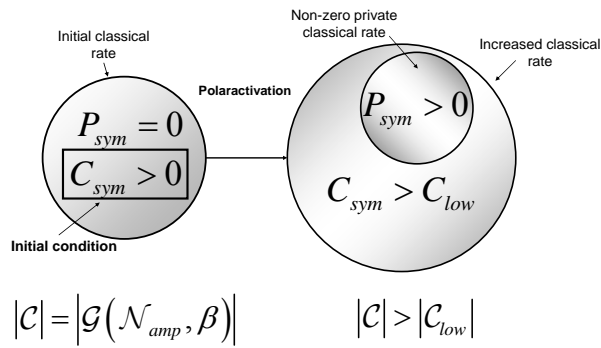

**Figure 6.** The increase in the rate of classical communication means that in the polarized channel structure the length of codewords that can transmit classical information is increased (the bits that convey non-private classical information in the *n*-length codeword form an (*n-l*)-length codeword). This increase in the codeword set makes possible to define new polar codewords that can transmit *l* private classical bits.

**Remark 3.** *The results on the connection of sets $\mathcal{C} = \mathcal{G}(\mathcal{N}_{amp}, \beta)$ and $S_{in}$ hold for both degradable and non-degradable channels, however the rate of the maximally achievable private classical communication will differ.*

In terms of these codeword sets, Corollary 2 and 3 can be revised as follows.

**Revision of Corollary 2.** *The lower bound on $|S_{in}|$ is determined by $|\mathcal{C}_{low}|$.*

**Revision of Corollary 3.** *The possible set of polaractivable channels is also determined by the distance $\Delta$ between the $|\mathcal{C}| = |\mathcal{G}(\mathcal{N}_{amp}, \beta)|$ and $|\mathcal{C}_{low}|$ the initial phase:*
$\Delta = |\mathcal{C}_{low}| - |\mathcal{G}(\mathcal{N}_{amp}, \beta)|$, *where* $|\mathcal{C}_{low}| > |\mathcal{C}|$.

**Corollary 4.** *As it also follows from the revised corollaries, the polaractivation of the private classical capacity domain is a natural consequence of the channel polarization effect in quantum systems. While in case of the classical polarization effect [1], only the rate of classical communication can be increased over a classical channel N, in case of a quantum channel $\mathcal{N}$ the improvement in $|\mathcal{C}| = |\mathcal{G}(\mathcal{N}_{amp}, \beta)|$ makes possible to reach a hidden subset of $\mathcal{C}$: set $S_{in}$.*

As it clearly follows from (69) and (70), $S_{in}$ is, indeed, a subset of $\mathcal{C}$: $S_{in} \subseteq \mathcal{C} = (\mathcal{G}(\mathcal{N}_{amp}, \beta) \cap \mathcal{G}(\mathcal{N}_{phase}, \beta)) \subseteq \mathcal{G}(\mathcal{N}_{amp}, \beta)$, and the opening of the subset of $S_{in}$ is determined only by the relation $|\mathcal{C}_{low}| > |\mathcal{C}|$. To prove the polaractivation of private classical capacity we have to show that by using quantum polar codes, the transition of $S_{in} = \varnothing$ into $S_{in} \neq \varnothing$ can be achieved. The valuable indexes of input message $A$ transmit the $l$-length private message. Eve will receive only garbage bits in the remaining $n - l$ bits of $A$, since the $(n\text{-}l)$-length codewords contain classical correlation, only (From the viewpoint of Eve these bits are random). From Alice's input message $M$, her $\mathcal{E}$ encoder creates a $n$-length message $A$. If $S_{in} \neq \varnothing$, then private communication is possible between Alice and Bob, and $l$ bits from the $A$ input message of $\mathcal{N}$ will be a codeword from the set $S_{in}$, denoted by $s_{in} \in S_{in}$. From the channel output message $B$, Bob's decoder $\mathcal{D}$ constructs the decoded private message $M'$. Using $p_{Eve}$ Eve's error probability and positive parameters $\gamma > 0$ and $\lambda > 1$ for codeword set $\mathcal{P}_1 \cup \mathcal{P}_2$:

$$|\mathcal{P}_1 \cup \mathcal{P}_2| \geq n\left(1 - p_{Eve} - \frac{1}{n^\lambda}\gamma\right). \tag{80}$$

From this result, for the set $S_{in} = \mathcal{G}(\mathcal{N}_{amp}, \beta) \cap \mathcal{G}(\mathcal{N}_{phase}, \beta)$:

$$|S_{in}| = n \cdot p_{Eve} + \frac{1}{n^{1-\lambda}}\gamma, \tag{81}$$

and

$$\begin{aligned}H(M) &= |\mathcal{G}(\mathcal{N}_{amp},\beta) \cap \mathcal{G}(\mathcal{N}_{phase},\beta)| \\ &= |S_{in}| = n \cdot p_{Eve} + \frac{1}{n^{1-\lambda}}\gamma.\end{aligned} \tag{82}$$

As depicted in Fig. 7, the polar codewords that convey the private classical information are generated by the encoder device. The private bits can be transmitted if and only if the polarization effect is achieved, which requires long codewords. The polarization effect increases the maximally achievable rate of classical communication over the polarized channel structure which makes possible to open the hidden private capacity-domain and the private bits can be embedded into the transmitted polar codeword.

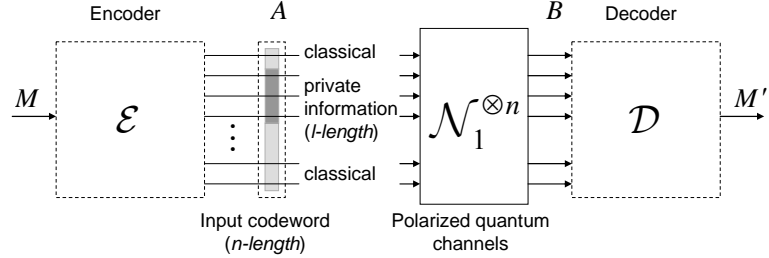

**Figure 7.** The quantum polar coding-based quantum private communication system. Alice constructs an $n$-length input codeword $A$, which consists of the $(n\text{-}l)$-length classical codeword and the $l$-length private codeword. The polaractivation increases $n$ and $(n\text{-}l)$, to make possible to transmission of the $l$-length private codeword.

Using $|S_{in}|$, it follows that the following private capacity can be achieved between Alice and Bob, assuming $n$ channel uses:

$$P_{sym}(\mathcal{N}_{Bob}) \geq C_{Bob} - \frac{1}{n^{\lambda}}\gamma, \tag{83}$$

and, this result is guaranteed by those codewords for which the following conditions are satisfied [1]. For the $l$ valuable bits from the given codeword $s_{in}$ selected from $|S_{in}|$, a $seq_n^i \leq C_{Bob} \cdot e^{n^\beta}, for\ i \in [n]$ sequence is generated using the initial condition $seq_1^{(1)} = p_{Eve}$:

$$seq_{2l}^{(2i-1)} = 2seq_{l}^{(i)} - \left(seq_{l}^{(i)}\right)^2, \text{ for } i \in [l] \\ seq_{2l}^{(2i)} = \left(seq_{l}^{(i)}\right)^2, \text{ for } i \in [l]. \tag{84}$$

For this sequence, as $n \to \infty$

$$H(M|E) \to H(M), \tag{85}$$

and, as a corollary, (81) and (82) are trivially satisfied, which concludes the proof on the achievability of codeword $s_{in}$ from the set $S_{in}$. The sets $\mathcal{P}_1$ and $\mathcal{P}_2$ are disjoint [1], [7], [16], [24], thus

$$|\mathcal{P}_1 \cup \mathcal{P}_2| = |\mathcal{P}_1| + |\mathcal{P}_2|, \tag{86}$$

and since Alice and Bob agreed to use only the set $\mathcal{P}_1$ for the transmission of the frozen bits (and only set $\mathcal{G}(\mathcal{N}_{amp}, \beta)$ used for classical communication), it leads to

$$\lim_{n \to \infty} \frac{1}{n} |\mathcal{P}_2| = 0, \tag{87}$$

and $|\mathcal{P}_2 \cap \mathcal{G}(\mathcal{N}_{amp}, \beta)| = 0$ with $|\mathcal{B}(\mathcal{N}_{amp}, \beta) \cap \mathcal{G}(\mathcal{N}_{amp}, \beta)| = 0$, which follows from the fact that $\mathcal{P}_2 \subseteq \mathcal{B}(\mathcal{N}_{amp}, \beta)$, where $\mathcal{B}(\mathcal{N}_{amp}, \beta) = [n] \setminus \mathcal{G}(\mathcal{N}_{amp}, \beta)$ and $|\mathcal{P}_2 \cap (S_{in} \cup \mathcal{B})| = 0$ along with $\mathcal{P}_1 \cup \mathcal{P}_2 \subseteq \mathcal{G}(\mathcal{N}_{amp}, \beta)$. For the proposed scheme, $\mathcal{P}_1 \subseteq \mathcal{G}(\mathcal{N}_{amp}, \beta)$, and

$$|\mathcal{G}(\mathcal{N}_{amp}, \beta)| = [n], \tag{88}$$

and

$$([n] \setminus (\mathcal{P}_1 \cup \mathcal{P}_2)) \subseteq \left(\mathcal{G}(\mathcal{N}_{amp}, \beta) \cap \mathcal{G}(\mathcal{N}_{phase}, \beta)\right) \\ \cup \left(\mathcal{B}(\mathcal{N}_{amp}, \beta) \cap \mathcal{G}(\mathcal{N}_{phase}, \beta)\right) \tag{89}$$

are satisfied, i.e., the empty set of private input codewords is transformed into a non-empty set

$$S_{in} = \mathcal{G}(\mathcal{N}_{amp}, \beta) \cap \mathcal{G}(\mathcal{N}_{phase}, \beta) \neq \varnothing, \tag{90}$$

which proves that if there exists a non-empty set $S_{in}$, then the polaractivation of private classical capacity of these channels can be achieved, which concludes the proof of Theorem 1.

∎

The proposed results on the achievable rate of secret private communication assuming a degradable channel between Alice and Bob are summarized in Theorem 2.

**Theorem 2.** *The symmetric private classical capacity of any degradable quantum channels for which the conditions of Section 2.5 are satisfied can be polaractivated.*

*Proof.* Assuming a *degraded* channel $\mathcal{N}_{Eve}$ [24], the following $P_{sym}$ symmetric private classical capacity can be achieved over the quantum channel $\mathcal{N}_{Bob}$:

$$P_{sym} = \lim_{n \to \infty} \frac{1}{n} \left( |S_{in}| \right)$$
$$= \lim_{n \to \infty} \frac{1}{n} \left| \mathcal{G}(\mathcal{N}_{amp}, \beta) \cap \mathcal{G}(\mathcal{N}_{phase}, \beta) \right|. \tag{91}$$

First, we give the proof of $P_{sym}$, then for the rate $R_{sym}$. Assuming $\beta < 0.5$,

$$C(\mathcal{N}_{Bob}) = \lim_{n \to \infty} \frac{1}{n} \left| \mathcal{G}(\mathcal{N}_{amp}, \beta) \right|, \tag{92}$$

$$C(\mathcal{N}_{Eve}) = \lim_{n \to \infty} \frac{1}{n} \left( |\mathcal{P}_1| + |\mathcal{P}_2| \right). \tag{93}$$

Combing this result with (87), we get

$$C(\mathcal{N}_{Eve}) = \lim_{n \to \infty} \frac{1}{n} \left( |\mathcal{P}_1| \right) \tag{94}$$

and

$$P(\mathcal{N}_{Bob}) = C(\mathcal{N}_{Bob}) - C(\mathcal{N}_{Eve})$$
$$= \lim_{n \to \infty} \frac{1}{n} \left| \mathcal{G}(\mathcal{N}_{amp}, \beta) - |\mathcal{P}_1| \right|. \tag{95}$$

The result obtained in (95) can be rewritten as follows:

$$P(\mathcal{N}_{Bob}) = \lim_{n \to \infty} \frac{1}{n} \left( |S_{in}| \cup |\mathcal{P}_2| \right) = \lim_{n \to \infty} \frac{1}{n} \left( |S_{in}| \right). \tag{96}$$

From the polar encoding scheme, it follows that

$$\sqrt{F(S_{in})} < 2^{-n^\beta}, \tag{97}$$

and for the fidelity parameters of $\mathcal{P}_1$:

$$\sqrt{F(\mathcal{P}_1)} \geq 1 - 2^{-n^\beta}. \tag{98}$$

If (97) and (98) are satisfied, then $\mathcal{P}_1 \cap \mathcal{B} \neq \varnothing$. Using the sets as defined in (70) and (73), it follows that [17], [24], [32]

$$\begin{aligned} \left(\mathcal{B}(\mathcal{N}_{amp}, \beta) \cap \mathcal{B}(\mathcal{N}_{phase}, \beta)\right) \subseteq \\ \left(S_{in} \cup \left(\mathcal{B}(\mathcal{N}_{amp}, \beta) \cap \mathcal{B}(\mathcal{N}_{phase}, \beta)\right)\right) = \varnothing. \end{aligned} \tag{99}$$

After some steps of reordering, we get that

$$\begin{aligned} \left(\mathcal{G}(\mathcal{N}_{amp}, \beta) \cap \mathcal{B}(\mathcal{N}_{phase}, \beta)\right) \\ \cap \left(\mathcal{B}(\mathcal{N}_{amp}, \beta) \cap \mathcal{B}(\mathcal{N}_{phase}, \beta)\right) \subseteq \\ \left(\mathcal{G}(\mathcal{N}_{amp}, \beta) \cap \mathcal{B}(\mathcal{N}_{phase}, \beta)\right) \\ \cap \left(S_{in} \cup \left(\mathcal{B}(\mathcal{N}_{amp}, \beta) \cap \mathcal{B}(\mathcal{N}_{phase}, \beta)\right)\right) = \varnothing, \end{aligned} \tag{100}$$

$$\mathcal{P}_1 \cap (S_{in} \cup \mathcal{P}_2) = \varnothing. \tag{101}$$

This result also means that the constructed codeword sets $S_{in}$, $\mathcal{P}_1$, $\mathcal{P}_2$, and $\mathcal{B}$ are disjoint sets with relation $|S_{in} \cup \mathcal{P}_1 \cup \mathcal{P}_2| = n$. Since Eve's channel is degraded [17], [24], [32],

$$\lim_{n \to \infty} \frac{1}{n} |\mathcal{B}| = 0, \tag{102}$$

which concludes our proof on $P_{sym}(\mathcal{N})$ for a degradable quantum channel:

$$P_{sym}(\mathcal{N}) = \lim_{n \to \infty} \frac{1}{n} |S_{in}|. \tag{103}$$

As follows, the maximal rate of private classical communication is:

$$\begin{aligned} P_{sym}(\mathcal{N}) = \max R_{sym} &= \lim_{n \to \infty} \frac{1}{n} \left(|S_{in}|\right) \\ &= \lim_{n \to \infty} \frac{1}{n} \left(|\mathcal{G}(\mathcal{N}_{amp}, \beta) \cap \mathcal{G}(\mathcal{N}_{phase}, \beta)|\right). \end{aligned} \tag{104}$$

In other words, if $\mathcal{N}_{Eve}$ is a degraded channel, then achievable codewords are

$$\left|\mathcal{G}(\mathcal{N}_{amp}, \beta) \cap \mathcal{G}(\mathcal{N}_{phase}, \beta)\right|, \tag{105}$$

from which the proof of (91) is concluded. These results conclude that for the non-empty sets $|S_{in}|$ the private classical capacity will be positive which concludes the proof on the polaractivation.

∎

**Theorem 3.** *The symmetric private classical capacity of any non-degradable quantum channel for which the conditions of Section 2.5 are satisfied can be polaractived.*

*Proof.* Assuming a degradable $\mathcal{N}_{Eve}$, for the sets $\mathcal{B}(\mathcal{N}_{amp}, \beta)$ and $\mathcal{P}_2$, the following relation holds for the private communication between Alice and Bob:

$$\begin{aligned} P_{sym}(\mathcal{N}) &= \lim_{n \to \infty} \frac{1}{n} \left| (|S_{in}| + |\mathcal{B}| - |\mathcal{B}(\mathcal{N}_{amp}, \beta)| + |\mathcal{P}_2|) \right| \\ &= \lim_{n \to \infty} \frac{1}{n} \left| \begin{array}{l} |\mathcal{G}(\mathcal{N}_{amp}, \beta) \cap \mathcal{G}(\mathcal{N}_{phase}, \beta)| \\ + |\mathcal{B}(\mathcal{N}_{amp}, \beta) \cap \mathcal{B}(\mathcal{N}_{phase}, \beta)| \\ - |\mathcal{B}(\mathcal{N}_{amp}, \beta)| + |\mathcal{B}(\mathcal{N}_{amp}, \beta) \cap \mathcal{G}(\mathcal{N}_{phase}, \beta)| \end{array} \right|, \end{aligned} \quad (106)$$

where for polar codeword set $\mathcal{P}_2$, $\lim_{n \to \infty} \frac{1}{n} |\mathcal{P}_2| = 0$, with error probability $p(M \neq M') = 2^{-n^\beta} + \sum_i F(\mathcal{N}_i)$, where $2^{-n^\beta}$ is the upper bound on set $\mathcal{G}(\mathcal{N}_{amp}, \beta)$ and $F$ is the fidelity of the $i$-th logical channel $\mathcal{N}_i$. Similar to the degraded case, the following set of polar codewords can be used for private classical communication:

$$\left| \mathcal{G}(\mathcal{N}_{amp}, \beta) \cap \mathcal{G}(\mathcal{N}_{phase}, \beta) \right|. \quad (107)$$

however, the decoding requires pre-shared entanglement between the frozen bits, i.e., $|\mathcal{B}| > 0$. As follows, if Eve's channel $\mathcal{N}_{Eve}$ is a non-degraded channel, then

$$P_{sym}(\mathcal{N}) = \lim_{n \to \infty} \frac{1}{n} \left| (|S_{in}| - |\mathcal{B}|) \right|. \quad (108)$$

These results show that positive private classical capacity can be achieved by the proposed polaractivation encoding scheme, over non-degradable quantum channels [17], [24], [32]. Next, we prove that the polaractivation of $P_{sym}$ can be achieved by using input codewords $S_{in}$ and assuming $n$ channel uses of the same non-degradable quantum channel $\mathcal{N}$. It follows that

$$\begin{aligned} P_{sym}(\mathcal{N}) &\geq P_{sym}^{(1)}(\mathcal{N}), \\ P_{sym}(\mathcal{N}) &\geq I(M:B) - p_{Eve} - \varepsilon, \end{aligned} \quad (109)$$

along with

$$P_{sym}(\mathcal{N}) \leq \frac{1}{n}|S_{in}| + p_{Eve} + \varepsilon, \tag{110}$$

where $\varepsilon > 0$. It is enough to show that there exists an input codeword $s_{in}$ in $S_{in}$, for which (109) and (110) are satisfied. Assuming $n \to \infty$ and $\varepsilon \to 0$, the following result holds for the given codewords $s_{in} \in S_{in}$, $b_{in} \in \mathcal{B}$ and $p_{in} \in \mathcal{P}_1 \cup \mathcal{P}_2$ [7]:

$$\frac{1}{n}I(p_{in}, s_{in} : b_{in}, E) \leq \frac{1}{n}|\mathcal{P}_1 \cup \mathcal{P}_1| + p_{Eve} \tag{111}$$

and using that

$$\frac{1}{n}H(s_{in}|E, b_{in}) + \varepsilon + \frac{1}{n}I(p_{in}, s_{in} : b_{in}, E) \geq \frac{1}{n}|S_{in}| + \frac{1}{n}|\mathcal{P}_1 \cup \mathcal{P}_2|, \tag{112}$$

where

$$\frac{1}{n}I(p_{in}, s_{in} : b_{in}, E) \to \frac{1}{n}|\mathcal{P}_1 \cup \mathcal{P}_2|. \tag{113}$$

From this result, it also follows that

$$\frac{1}{n}H(s_{in}|E, b_{in}) \geq \frac{1}{n}|S_{in}| - \varepsilon - p_{Eve}, \tag{114}$$

where $\varepsilon \to 0$ and $p_{Eve} \to 0$ along with $n \to \infty$, results in

$$\frac{1}{n}H(s_{in}|E, b_{in}) \geq \frac{1}{n}|S_{in}| \tag{115}$$

for the given input codeword $s_{in} \in S_{in}$. From $P_{sym}(\mathcal{N}) \leq \frac{1}{n}|S_{in}|$, and the input codeword sets constructed $S_{in}, \mathcal{B}, \mathcal{P}_1$, and $\mathcal{P}_2$ and from (113) along with (115), it follows that

$$\begin{aligned}P_{sym}(\mathcal{N}) &= \lim_{n \to \infty} \frac{1}{n}|(|S_{in}| - |\mathcal{B}|)| \\ &= \lim_{n \to \infty} \frac{1}{n}|(|\mathcal{G}(\mathcal{N}_{amp}, \beta)| + |\mathcal{G}(\mathcal{N}_{phase}, \beta)| - n)| \\ &= \lim_{n \to \infty} \left|\frac{|\mathcal{G}(\mathcal{N}_{amp}, \beta)| + |\mathcal{G}(\mathcal{N}_{phase}, \beta)|}{n} - 1\right|.\end{aligned} \tag{116}$$

To conclude the proof, finally we get

$$\frac{1}{n}|S_{in}| - \frac{1}{n}|\mathcal{B}| = \lim_{n\to\infty} \frac{1}{n}\big(I(A:B) - I(A:E)\big) \geq \tag{117}$$
$$P_{sym}^{(1)}(\mathcal{N}) = I(A:B) - I(A:E),$$

where

$$P_{sym}^{(1)}(\mathcal{N}) \leq |S_{in}|. \tag{118}$$

From (117) and (118), it follows that (109) and (110) hold, which concludes the first part of the proof. Next, we show that

$$P_{sym}(\mathcal{N}) \geq \lim_{n\to\infty} \frac{1}{n}|S_{in}|. \tag{119}$$

From the basic properties of the polaractivation encoding scheme, for the fidelity parameters [1]

$$\sqrt{F(\mathcal{N}_i)} < 2^{-n^\beta}, \text{ for all } i \in S_{in}. \tag{120}$$

For $|S_{in}|$, the inequality

$$\frac{1}{n}|S_{in}| \leq \frac{1}{n}\big((n(C_{Bob} - \varepsilon)) - 1\big) - \frac{1}{n}\big((n(C_{Eve} - \varepsilon)) - 1\big)$$
$$= \lim_{n\to\infty} \frac{1}{n}|S_{in}| \leq (C_{Bob}) - (C_{Eve}) - 2\varepsilon - \frac{2}{n} = P_{sym}(\mathcal{N}) - 2\varepsilon - \frac{2}{n} \tag{121}$$

also follows. From (121), we can write for the symmetric private capacity that

$$P_{sym}(\mathcal{N}) \leq \frac{1}{n}|S_{in}| + 2\varepsilon + \frac{2}{n}. \tag{122}$$

Furthermore, if $n \to \infty$ and $\mathcal{E} \to 0$, then

$$P_{sym} \geq \frac{1}{n}|S_{in}|. \tag{123}$$

From (123)

$$\lim_{n\to\infty} \frac{1}{n}\left|\left(|S_{in}| - |S_{bad}|\right) + \left|\begin{matrix}(\mathcal{G}(\mathcal{N}_{amp},\beta) \cap \mathcal{B}(\mathcal{N}_{phase},\beta)) \\ \cup (\mathcal{B}(\mathcal{N}_{amp},\beta) \cap \mathcal{G}(\mathcal{N}_{phase},\beta))\end{matrix}\right|\right| \leq \frac{1}{n}|S_{in}| \tag{124}$$

is also satisfied. Assuming that the error probability of Eve's channel $\mathcal{N}_{Eve}$ is $p_{Eve}$, the following lower bound can be given [7] for the $H(M|E)$ conditional entropy:

$$H(M|E) \geq n \cdot p_{Eve}\left(1 - c2^{-n^\beta}\right), \tag{125}$$

where $c$ is a positive constant. From this result, for $I(M:B)$ the following lower bound holds [7], [17], [24], [32]:

$$I(M:B) \geq p_{Eve}\left(1 - c2^{-n^{\beta}}\right), \quad (126)$$

and

$$H(M|B,E) + H(B|E) - H(B|M,E) \geq n \cdot p_{Eve}\left(1 - c2^{-n^{\beta}}\right). \quad (127),$$

which concludes our proof on (119) and (122).

∎

**Corollary 5.** *For a non-degradable quantum channel the polaractivated private capacity (the maximal rate of private classical communication) is*

$$P_{sym} = \max R_{sym} = \lim_{n\to\infty} \frac{1}{n}\left|\left(|S_{in}| - |S_{bad}| + \left|\begin{array}{l}\left(\mathcal{G}(\mathcal{N}_{amp},\beta) \cap \mathcal{B}(\mathcal{N}_{phase},\beta)\right) \\ \cup \left(\mathcal{B}(\mathcal{N}_{amp},\beta) \cap \mathcal{G}(\mathcal{N}_{phase},\beta)\right)\end{array}\right|\right)\right|. \quad (128)$$

From the codewords set construction scheme [17], [24], [32] follows that

$$\begin{aligned}\lim_{n\to\infty} \frac{1}{n}\left|\left(|S_{bad}| - \left|\begin{array}{l}\left(\mathcal{G}(\mathcal{N}_{amp},\beta) \cap \mathcal{B}(\mathcal{N}_{phase},\beta)\right) \\ \cup \left(\mathcal{B}(\mathcal{N}_{amp},\beta) \cap \mathcal{G}(\mathcal{N}_{phase},\beta)\right)\end{array}\right|\right)\right| \\ = \lim_{n\to\infty} \frac{1}{n}\left|\left(\left|\mathcal{B}(\mathcal{N}_{amp},\beta) \cap \mathcal{B}(\mathcal{N}_{phase},\beta)\right|\right)\right|,\end{aligned} \quad (129)$$

with

$$\begin{aligned}\left|\mathcal{G}(\mathcal{N}_{amp},\beta) \cap \mathcal{G}(\mathcal{N}_{phase},\beta)\right| = \\ \left|\mathcal{G}(\mathcal{N}_{amp},\beta)\right| + \left|\mathcal{G}(\mathcal{N}_{phase},\beta)\right| - \left|\mathcal{G}(\mathcal{N}_{amp},\beta) \cup \mathcal{G}(\mathcal{N}_{amp},\beta)\right|,\end{aligned} \quad (130)$$

which leads us to symmetric private classical capacity:

$$\begin{aligned}P_{sym} &= \lim_{n\to\infty} \frac{1}{n}\left|\left(\left|\mathcal{G}(\mathcal{N}_{amp},\beta)\right| + \left|\mathcal{G}(\mathcal{N}_{phase},\beta)\right| - n\right)\right| \\ &= \lim_{n\to\infty}\left|\left(\frac{\left|\mathcal{G}(\mathcal{N}_{amp},\beta)\right| + \left|\mathcal{G}(\mathcal{N}_{phase},\beta)\right|}{n} - 1\right)\right|.\end{aligned} \quad (131)$$

From the proposed encoding scheme follows, that for the positive $P_{sym}$ symmetric capacity there exits the codeword set $S_{in} \neq \varnothing$, and the theorem is proven for any non-degradable quantum channels for which the conditions of Section 2.5 are satisfied.

*5.1. Brief Conclusion*

From Theorems 2 and 3, the polaractivation of the symmetric private classical capacity of *arbitrary* degradable and non-degradable quantum channels (for which the conditions of Section 2.5 are satisfied) is proven.

As summarized in Fig. 8, the polaractivation will result in the non-empty set $S_{in} \neq \varnothing$, and the channel will be able to transmit classical information privately.

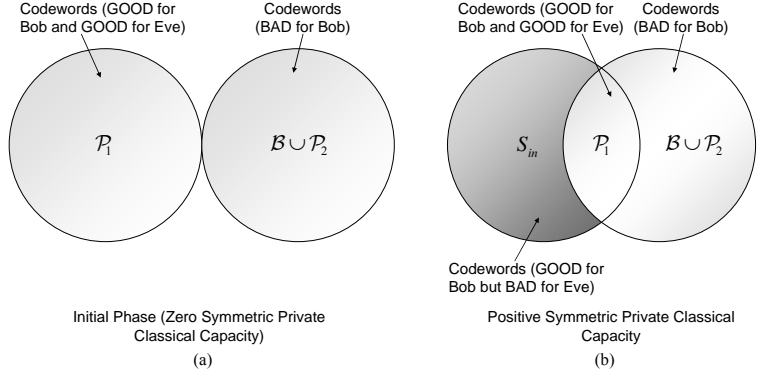

**Figure 8 (a)**. The brief summarization of the proposed polaractivation scheme. In the initial phase, the input channels cannot transmit classical information privately. **(b)** The polaractivation of symmetric private classical capacity makes it possible to construct codewords capable of transmitting private classical information between Alice and Bob.

# 6. Numerical Evidences

In this section we illustrate the theorems with numerical examples on the polaractivation. The results will be demonstrated for qubit Pauli channels with error probability $p = p_x + p_y + p_z$.

*6.1. Polaractivation of Private Classical Capacity for Arbitrary Channels by the Constructed Codewords*

In this section we demonstrate the theoretical results and the proposed theorems by a numerical evidence. Here, we analyze the properties of our quantum polar-encoding scheme using a numerical example. We assume that for the quantum channel $\mathcal{N}_{Bob}$ the conditions of Section 2.5 are satisfied. We demonstrate the results for a non-degraded and a degraded eavesdropper channel $\mathcal{N}_{Eve}$. For the non-degraded case, $p_{Bob}(M \neq M) > p_{Eve}(M \neq M)$, while for the degraded case, $p_{Bob}(M \neq M) < p_{Eve}(M \neq M)$. The achievable $C_{sym}$ symmetric classical and $P_{sym}$ symmetric private classical capacities are expressed by the constructed codeword sets, $S_{in}, \mathcal{B}, \mathcal{P}_1$, and $\mathcal{P}_2$. In our analysis, the number $n$ of channel uses of $\mathcal{N}$ was chosen to $2^{15}$ and $2^{25}$, which are values that were determined in accordance with [1]. For the degraded eavesdropper channel $\mathcal{N}_{Eve}$, $\lim_{n \to \infty} \frac{1}{n} |\mathcal{B}| = 0$.

The error rates $p_{Bob}(M \neq M)$, $p_{Eve}(M \neq M)$ of Bob and Eve for the non-degraded and the degraded cases were chosen according to the relation $F_x + F_z < 1$ and $F_x + F_z \geq 1$ of the fidelities, where $F_z$ and $F_x$ related to the error of the amplitude and phase transmission over the polarized channel structure $\mathcal{N}^{\otimes n}$, and defined as follows:

$$F_z(\mathcal{N}) = 2\sqrt{p_z(1 - p_z)} \text{ and } F_x(\mathcal{N}) = 2\sqrt{p_x(1 - p_x)}, \tag{132}$$

where $p_z$ and $p_x$ are the amplitude and phase error probabilities occurring independently on the noisy quantum channel $\mathcal{N}$.

Fig. 9 illustrates the $P_{sym}$ achievable private classical capacity for a non-degraded and a degraded eavesdropper as the function of the Eve's information. The private classical capacity is expressed as $\frac{||S_{in}| - |\mathcal{B}||}{n}$ for the non-degraded case, and as $\frac{|S_{in}|}{n}$ for the degraded case. The $x$-axis represents Eve's channel capacity, expressed as $\frac{|\mathcal{P}_1| + |\mathcal{P}_2|}{n}$ and $\frac{|\mathcal{P}_1|}{n}$ for the degraded case, while the $y$-axis represents symmetric private classical capacity $P_{sym}$. The number $n$ of channel uses was chosen to $2^{15}$ and $2^{25}$, respectively.

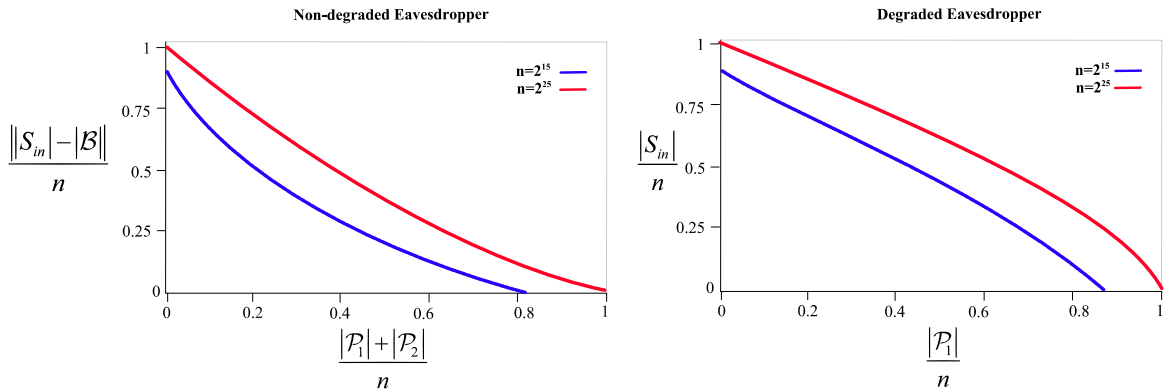

**Figure 9.** The achievable private classical communication rate expressed as Bob's valuable codewords in the function of Eve's valuable codewords for a non-degraded eavesdropper **(a)** and for a degraded eavesdropper **(b)**.

For a degraded channel $\mathcal{N}_{Eve}$, the rate $R$ of the private classical communication can be increased, and $\frac{|S_{in}|}{n} > \frac{\||S_{in}|-|\mathcal{B}|\|}{n}$ will hold, since $\frac{|\mathcal{P}_1|}{n} < \frac{|\mathcal{P}_1|+|\mathcal{P}_2|}{n}$. The maximally achievable private classical capacity $P_{sym}$ can be achieved, and higher rates of $R$ can be approached. For the non-degraded case, if $n = 2^{15}$, the $\frac{\||S_{in}|-|\mathcal{B}|\|}{n}$ converges to 0 as $\frac{|\mathcal{P}_1|+|\mathcal{P}_2|}{n} \to 0.8$, while for a larger number of $n$, (i.e., for $n = 2^{25}$), these results occur only if $\frac{|\mathcal{P}_1|+|\mathcal{P}_2|}{n} \to 1$ holds. Similar results were obtained for a degraded channel $\mathcal{N}_{Eve}$, for $n = 2^{15}$, $\frac{|S_{in}|}{n} \to 0$ if $\frac{|\mathcal{P}_1|}{n} \to 0.85$ and $\frac{|\mathcal{P}_1|}{n} \to 1$, since for the degraded case, $\lim_{n \to \infty} \frac{1}{n}|\mathcal{B}| = 0$.

**Theorem 4.** *The polaractivation of symmetric private classical capacity depends on the amount of available symmetric classical capacity $C_{sym.}$, which can be achieved by the proposed polar coding technique.*

*Proof.* In Fig. 10(a) and Fig. 10(b) the relation of the fidelity parameter $F = F_x + F_z$ (where $B_x$ is the parameter of the *phase* transmission, and $F_z$ is related to the *amplitude* transmission) and the sets of $\frac{\||S_{in}|-|\mathcal{B}|+|\mathcal{P}_1|+|\mathcal{P}_2|\|}{n}$ and of $\frac{\||S_{in}|-|\mathcal{B}|\|}{n}$ are shown. In Fig. 10(a), the *x*-axis represents $C_{sym}$, while in Fig. 10(b) the private classical capacity $P_{sym}$ is depicted. The *y*-axis illustrates the sum of the fidelity parameters $F_x$ and $F_z$ of the phase and amplitude transmission. In the case of Fig. 10(a), $n = 2^{15}$ and $n = 2^{25}$, while for Fig. 10(b), $n = 2^{25}$. The set $\frac{\||S_{in}|-|\mathcal{B}|+|\mathcal{P}_1|+|\mathcal{P}_2|\|}{n}$ represents the $C_{sym}$ classical symmetric capacity between Alice and Bob, while the set $\frac{\||S_{in}|-|\mathcal{B}|\|}{n}$ refers to the $P_{sym}$ private classical capacity, assuming a non-degradable channel $\mathcal{N}$.

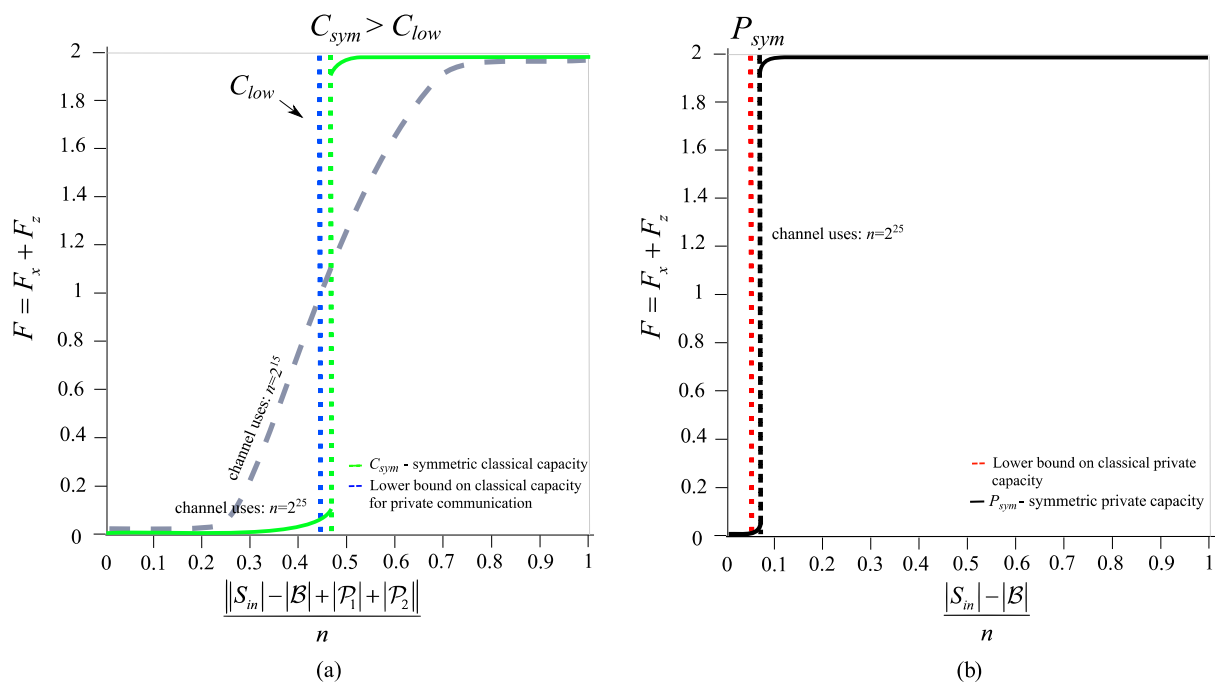

**Figure 10 (a)**: The ratio of the fidelity parameters of "good" and "bad" channels as the function of the achievable symmetric classical capacity expressed by the codewords of Alice and Bob. As the set converges to the critical lower bound $C_{low}$, the symmetric private classical capacity will becomes positive, otherwise positive private capacity is not possible. **(b)**: The ratio of the fidelity parameters of "good" and "bad" channels as the function of the achievable symmetric *private* classical capacity expressed by the private codewords of Alice and Bob. For high enough number *n* of channel uses, the $P_{sym} > 0$ critical lower bound can be exceeded, as depicted by the black dashed line, which indicates that $C_{sym} \geq C_{low}$ is also satisfied.

In Fig. 10(a), the connection between the number $n$ of channel uses and $C_{sym}$, the achievable symmetric classical capacity is depicted. For the relatively low number of channel uses $n = 2^{15}$, the fraction of "good" channel indexes does not reach the critical symmetric classical capacity $C_{low}$, i.e., the fraction of the fidelity parameters of the "good" channels converge to a $C_{sym}$, where $C_{sym} < C_{low}$. The critical lower bound $C_{low}$ cannot exceed for $n = 2^{15}$; however, for $n = 2^{25}$, $C_{sym} \geq C_{low}$ can be achieved, which makes possible the private classical communication between Alice and Bob. On the other hand, for $n = 2^{15}$, $C_{sym} < C_{low}$ and positive private capacity cannot be achieved.

In Fig. 10(b). the achievable private capacity is shown for $n = 2^{25}$. From these results, it can be obtained, that $n = 2^{25}$ is a good choice for Alice, since in this case $P_{sym} > 0$ and the critical lower bound is also exceeded, i.e., $C_{sym} > C_{low}$, which is depicted by the black dashed line. For $n = 2^{25}$, the fraction of the fidelities $F = 0$ reaches and exceeds the critical $C_{low}$, which results in $P_{sym} > 0$. The critical lower bound on $P_{sym} > 0$ is depicted by the red dashed line. This lower bound is exceeded for $n = 2^{25}$. The numerical evidence indicates that achieving the private classical capacity is possible if, and only if, a $C_{low}$ critical lower bound in the $C_{sym}$ symmetric classical capacity is exceeded.

∎

The exact value of the critical lower bound $C_{low}$ can be determined from the fidelity parameters $F_x$ and $F_z$ of the channel indexes, as we will discuss next.

*6.2. Exact Lower Bound on the Symmetric Classical Capacity*

Our result on the connection of the polaractivation of the symmetric private classical capacity and the lower bound on the amount of the required symmetric classical capacity are summarized in Theorem 5.

**Theorem 5.** *The polaractivation of private classical capacity depends on the amount of available symmetric classical capacity $C_{sym.}$, which can be achieved by the polar coding technique. The transition of $C_{low.}(\mathcal{N}) \to P_{sym.}(\mathcal{N}) > 0$ can be achieved if and only if the relation $C_{sym.} < C_{low.}$ can be transformed into $C_{sym.} \geq C_{low.}$.*

*Proof:* Assuming the quantum channel $\mathcal{N}$, which can transmit amplitude and phase, the symmetric classical capacity can be expressed as follows [16], [17]:

$$C_{sym.}(\mathcal{N}) \geq \log_2 \frac{2}{1 + F_z(\mathcal{N})} + \log_2 \frac{2}{1 + F_x(\mathcal{N})}, \tag{133}$$

and

$$C_{sym.}(\mathcal{N}) \leq \sqrt{1 + F_z(\mathcal{N})^2} + \sqrt{1 + F_x(\mathcal{N})^2}. \tag{134}$$

In Fig. 11(a), the $F_z$ and $F_x$ of the amplitude and phase transmission are shown. The input channel is a quantum channel with independent amplitude and phase noise. In that region, the channel can be used to transmit private classical information. In Fig. 11(b), the achievable symmetric classical capacity $C_{sym.}$ and the symmetric private capacity $P_{sym.}$ are shown. Initially, the input channel $\mathcal{N}_1$ has $P_{sym.} = 0$, i.e., it cannot be used to send private classical information. On the other hand, as we discovered, the polaractivation of symmetric private capacity $P_{sym.}$ is possible by polar coding, but it requires the existence of an exact $C_{low}$ lower bound on the classical capacity $C_{sym.}$, i.e. if the transmission of private information is not possible (red area). If the classical capacity is above the lower bound $C_{low}$, then the channels have positive $P_{sym.}$ symmetric private capacity. Assuming $F_z > 0, F_x = 0$ or $F_z = 0, F_x > 0$, the lower bound on the symmetric classical capacity $C_{sym.}$, which is required for the positive symmetric private capacity $P_{sym.}$ is $C_{low} \geq 1 + 0.074$. In the initial phase, the channels are in the red area (i.e., have private capacity $P_{sym.} = 0$), in the end, the blue area of $C_{sym.}$ can be achieved, resulting in $P_{sym.} > 0$.

Using our encoding scheme of Fig 11(a), if $F_x + F_z \geq 1$, on the other hand, if $F_x + F_z < 1$, then it can classical information *privately.*

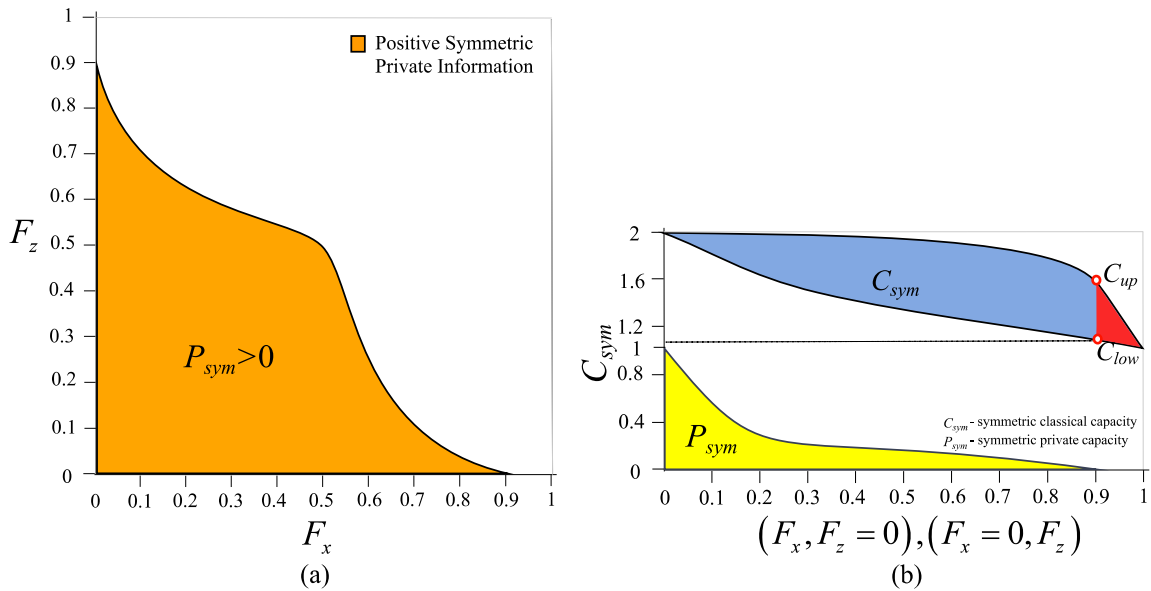

**Figure 11 (a)**. The fidelity parameters of the individual amplitude and phase-errors. The private classical capacity will be greater than zero only in this domain. **(b).** The achievable symmetric classical capacity and symmetric private classical capacity as the function of fidelity parameters. Individually, the channels are so noisy that they are in the red area; thus they cannot transmit any private classical information. The private capacity completely vanishes from $F_z = 0.9$ or $F_x = 0.9$ (besides $F_z = 0$ or $F_z = 0$).

In our case, individually, the channels are so noisy such that $F_x + F_z \geq 1$ (i.e., $P_{sym.} = 0$), but using the recursive channel construction scheme of the polar encoding scheme, the sum of the fidelity parameters starts to converge to zero (i.e., $F_x + F_z < 1$ will hold), which makes possible to use these noisy channels for the transmissions of private information and $P_{sym.} > 0$ can be achieved. Alice, using her amplitude and phase coding (see Fig. 11(a)), can transmit her private information to Bob, since for $F_x + F_z < 1$ the relation $1 - F_x - F_z \geq 0$ will always be true, which also determines $C_{low}$, the lower bound on $C_{sym.}$.

In Fig. 12(a), the achievable symmetric classical capacity $C_{sym.}$ and symmetric private capacity $P_{sym.}$ as the functions of the fidelity parameters $F_z$ and $F_x$ of the amplitude and phase trans-missions are shown. In Fig. 12(b), the critical area and the critical lower bound $C_{low.}$ on the classical symmetric capacity $C_{sym.}$ are shown. In the initial phase, the quantum channels are so noisy that they cannot transmit private information, (i.e., $P_{sym.} = 0$); however, they have some symmetric classical capacity, as shown in the red area. After the recursion steps of the channel iteration process (for details see [1], [2], [16], [17]) and the channel polarization effect are realized, and the fidelity parameters $F_z, F_x$ start to converge to zero, and for their sum and $F_z + F_x < 1$ will hold. (For the complete description of the iteration process, see Arikan's works [1], [2], [3].)

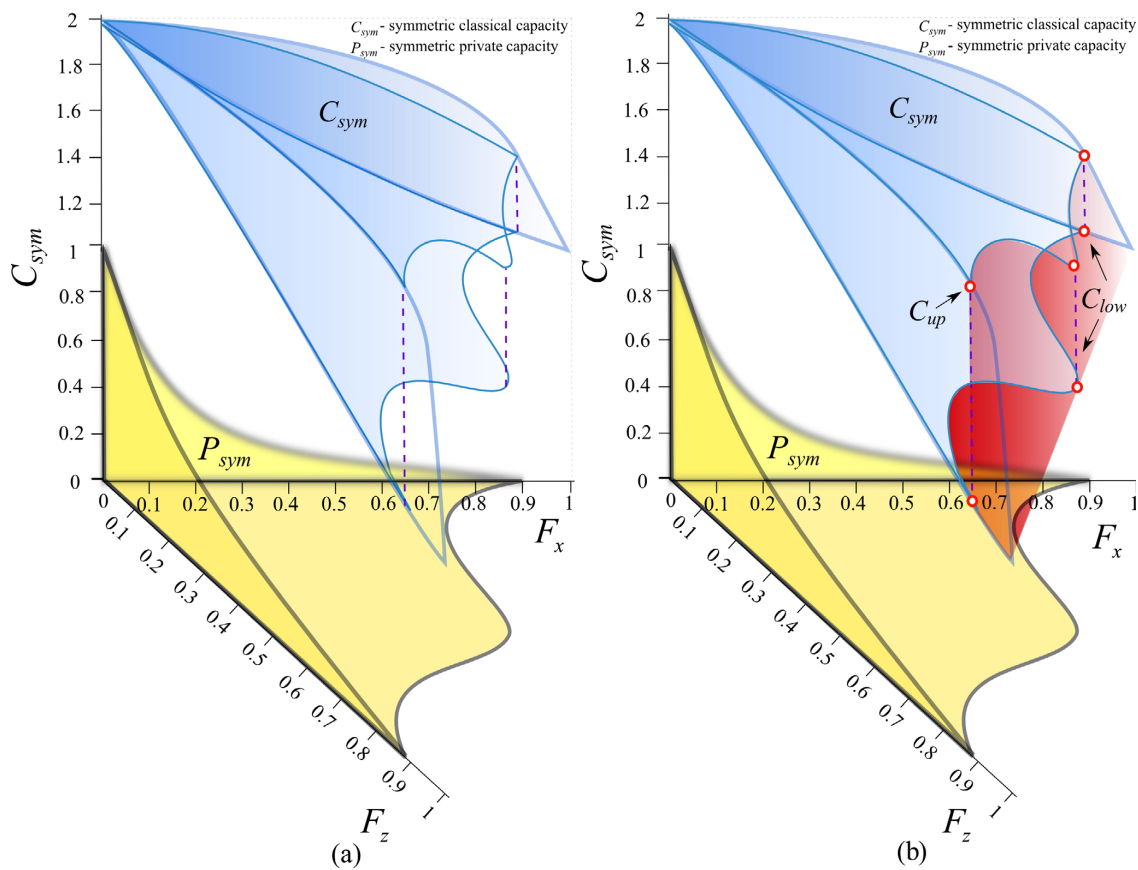

**Figure 12 (a)**. The achievable symmetric classical capacity (blue) and symmetric private classical capacity (yellow) as the functions of the fidelity parameters of the amplitude and phase transmissions of the noisy quantum channel. **(b)** The private communication is possible if and only if the polar codes can ensure the symmetric classical capacity above the critical lower bound. In the red area, the channels are so noisy that private communication is not possible. Initially, the input channels are in this area; however, with the help of polar coding the critical lower bound on the symmetric classical capacity can be exceeded, which makes possible private communication.

As we have confirmed, the amount of maximally polaractivated symmetric private capacity strictly depends on the level of maximally achievable symmetric classical capacity $C_{sym.}(\mathcal{N})$ of $\mathcal{N}$ for which $C_{sym.}(\mathcal{N}) \rightarrow P_{sym.}(\mathcal{N}) > 0$ is satisfied. (*Note:* The symmetric classical capacity is taken individually for the two individual channels which transmit the amplitude with $p_z$ error (with capacity $C_{sym.}^{ampl.}(\mathcal{N})$), and the phase with $p_x$ error probability (i.e., $C_{sym.}^{phase}(\mathcal{N})$).)

∎

*6.3. Error Probabilities*

In this section we illustrate how the error-probability of the transmission, and hence the amount of polaractivated private classical capacity depends on the number of channel uses.

**Theorem 6.** *The amount of polaractivated private classical capacity can be increased in the asymptotic setting.*

*Proof.* The private classical capacity $P_{sym.}$ can be polaractivated only in the asymptotic setting. The number of channel uses of $\mathcal{N}$ was $n = 2^{10}, 2^{15}$ and $2^{25}$. The ratio of the bad channels with '1' fidelity and good channels with '0' fidelity equals the private classical capacity $P_{sym.}(\mathcal{N}) > 0$.

In Fig. 13(a), the evolution the fidelities $F_z$ and $F_x$ of the amplitude and phase transmissions are shown, assuming a noisy quantum channel $\mathcal{N}_1$ assuming symmetric classical capacity $C_{sym.} = 0.47$ and initial fidelities $F_z = 0.7$, $F_x = 0.7$ (private communication is not possible in this domain, see Fig. 12).

In Fig. 13(b), the maximally achievable symmetric classical capacity as the function of the upper bound on the $p_b \in \left[2^{-22}, 2^{-2}\right]$ block error probability is shown, assuming $C_{low} = 2 \cdot 0.415 = 0.83$. Below the $C_{low}$ critical symmetric classical capacity, the channel cannot transmit private classical information. As it is also depicted, using different number of channel uses $n = 2^{10}, 2^{15}$ and $2^{25}$, the upper bounds on the $p_b$ block error probability for the polaractivated symmetric private capacity are also different. For higher numbers of channel uses, the $p_b$ upper bound on the error probability will be significantly lower.

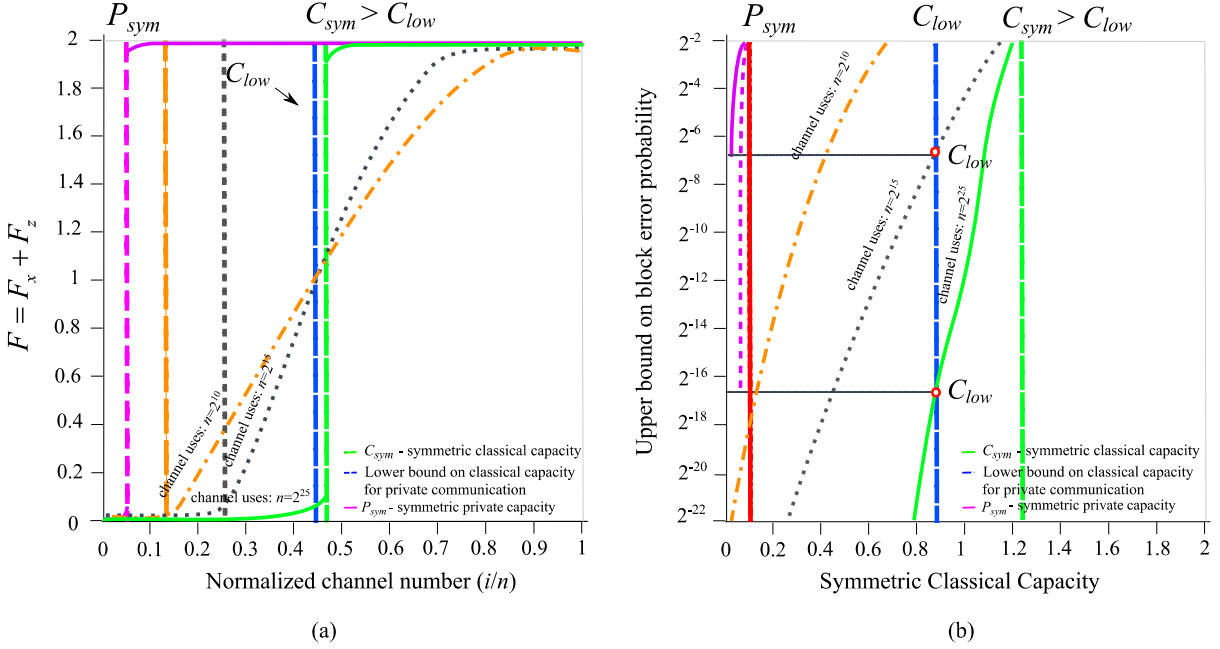

**Figure 13 (a).** The evolution of the fidelity parameters of the amplitude and phase transmissions of the channel. The achievable symmetric classical and symmetric private capacities are depicted by the green and red dashed lines. Due to the channel polarization, for a given set of channels, the fidelity parameter will be zero or one, and the ratio between the two sets is equal to the channel capacity. The polaractivation of symmetric private classical capacity requires the symmetric classical capacity to be above the critical lower bound. **(b).** The achievable classical capacity as the function of the upper bound on the block error probability, for various number $n$ of channel uses. The private capacity will be positive if $F_x + F_z < 1$, which can be achieved by the increasing number of channel uses.

∎

As an interesting conclusion from Figs. 13(a) and (b), the initial fidelity parameters can be improved by the increasing number $n$ of channel uses. Initially $F_x + F_z \geq 1$, while after the increasing of the number of channel uses, the required condition for the private communication (i.e., $F_x + F_z < 1$) could be achieved for the given values of $p_z$ and $p_x$, the amplitude and phase errors of the quantum channel.

# 7. Conclusions

In this paper, we introduced the term polaractivation for the opening of hidden private classical capacity region of quantum channels. The result is similar to the superactivation effect, however it is based on a fundamentally different theoretical background. Polaractivation is limited neither by any preliminary conditions on the maps of other channels involved in the joint channel structure and it also can be extended to the private classical capacity. Besides these advantages, it is limited by the rate of symmetric classical communication of the channel and the critical lower bound on it. We have also investigated that the critical lower bound in the rate of classical communication for the polaractivation can be exceeded by the quantum polar coding technique. In future work we will extend the proposed scheme for quantum channels which have no any strict requirements on the symmetric classical capacity and independent from the critical lower bound. We will also prove that the extended version of polaractivation can be used to transmit information over channels that have no hidden capacity-domains – the zero-capacity channels.

# Acknowledgements

The results discussed above are supported by the grant TAMOP-4.2.2.B-10/1--2010-0009 and COST Action MP1006.